\documentclass[sigconf,nonacm]{acmart}

\usepackage{lipsum}
\usepackage{float}
\usepackage{balance}
\usepackage{fontawesome5}

\usepackage{lscape} 
\usepackage{paralist}
\usepackage{subcaption}
\usepackage{xcolor,colortbl}
\definecolor{green}{rgb}{0.1,0.9,0.1}
\definecolor{red}{rgb}{0.9,0.1,0.1}

\usepackage[utf8]{inputenc}
\usepackage{hyperref}
\usepackage{graphicx}
\usepackage{fancyvrb}
\usepackage{array}
\usepackage{algpseudocode}
\usepackage{cprotect}
\usepackage{makecell}

\usepackage[linesnumbered,ruled,noend]{algorithm2e}

\usepackage{pifont}

\usepackage{comment}

\usepackage{enumitem}

\usepackage{adjustbox}
\usepackage{multirow}
\usepackage{makecell}
\usepackage{bm}

\usepackage{hhline}

\usepackage{graphicx}

\let\oldnl\nl
\newcommand{\nonl}{\renewcommand{\nl}{\let\nl\oldnl}}

\definecolor{bgred}{RGB}{255,210,205}
\definecolor{bgblue}{RGB}{210,220,255}
\definecolor{bgyellow}{RGB}{240,240,190}
\definecolor{bggrey}{RGB}{223,223,225}
\definecolor{purple}{RGB}{180,0,180}

\definecolor{hvygreen}{RGB}{50,200,0}
\definecolor{hvypink}{RGB}{102,0,51}

\newcolumntype{?}{!{\vrule width 2pt}}

    \newcommand{\annotation}[1]{[[[#1]]]}

    \newcommand{\amit}[1]{\textbf{\small\textcolor{blue}{\annotation{(Amit)~#1}}}{\typeout{#1}}}

    \newcommand{\oz}[1]{\textbf{\small\textcolor{cyan}{\annotation{(Oz)~#1}}}{\typeout{#1}}}

\makeatletter
\def\old@comma{,}
\catcode`\,=13
\def,{%
  \ifmmode%
    \old@comma\discretionary{}{}{}%
  \else%
    \old@comma%
  \fi%
}
\makeatother

    \newcommand{\system}{LINX}
    \newcommand{\lang}{LDX}
    \newcommand{\lquery}{Q_{X}}

    \newcommand{\spec}{s}
    \newcommand{\stree}{T_D}
    
    \newcommand{\e}[1]{\texttt{#1}}

    \newcommand{\slist}{\lquery}
    
    \newcommand{\nnodes}{Nodes(\lquery)}

    \newcommand{\conts}{Cont(\lquery)}
    \newcommand{\pluseq}{\mathrel{+}=}

        \newcommand{\basebin}{Binary Reward Only}
        \newcommand{\basedist}{Binary+Imm. Reward}
        \newcommand{\baseimm}{W/O Spec.Aware NN}

\begin{document}

\title{LINX: A Language Driven Generative System for Goal-Oriented Automated Data Exploration}

\author{Tavor Lipman}
\affiliation{%
  \institution{Tel Aviv University}
  \country{Israel}
}
\email{tavorlipman@mail.tau.ac.il}

\author{Tova Milo}
\affiliation{%
  \institution{Tel Aviv University}
  \country{Israel}
}
\email{milo@cs.tau.ac.il}

\author{Amit Somech}
\affiliation{%
  \institution{Bar-Ilan University}
  \country{Israel}
}
\email{somecha@cs.biu.ac.il}

\author{Tomer Wolfson}
\affiliation{%
  \institution{Tel Aviv University}
  \country{Israel}
}
\email{tomerwol@mail.tau.ac.il}

\author{Oz Zafar}
\affiliation{%
  \institution{Tel Aviv University}
  \country{Israel}
}
\email{ozzafar@mail.tau.ac.il}

\begin{abstract}

Data exploration is a challenging and time-consuming process in which users examine a dataset by iteratively employing a series of queries. While in some cases the user explores a new dataset to become familiar with it, more often, the exploration process is conducted with a specific analysis goal or question in mind.
To assist users in exploring a new dataset, Automated Data Exploration (ADE) systems have been devised in previous work. These systems aim to auto-generate a full exploration session, containing a sequence of queries that showcase interesting elements of the data. However, existing ADE systems are often constrained by a predefined objective function, thus always generating the same session for a given dataset. Therefore, their effectiveness in goal-oriented exploration, in which users need to answer specific questions about the data, are extremely limited.

To this end, this paper presents LINX, a generative system augmented with a natural language interface for goal-oriented ADE.
Given an input dataset and an analytical goal described in natural language, LINX generates a personalized exploratory session that is relevant to the user's goal.
LINX utilizes a Large Language Model (LLM) to interpret the input analysis goal, and then derive a set of specifications for the desired output exploration session. These specifications are then transferred to a novel, modular ADE engine based on Constrained Deep Reinforcement Learning (CDRL), which can adapt its output according to the specified instructions.

To validate LINX's effectiveness, we introduce a new benchmark dataset for goal-oriented exploration and conduct an extensive user study. Our analysis underscores LINX's superior capability in producing exploratory notebooks that are significantly more relevant and beneficial than those generated by existing solutions, including ChatGPT, goal-agnostic ADE, and commercial systems.

\end{abstract}

\maketitle


\section{introduction}
\label{sec:intro}

Data exploration is the process of examining a dataset by applying a sequence of queries, allowing the user to inspect different aspects of the data. Data exploration can be performed in one of two scenarios. The first, examining an unfamiliar dataset in order to understand its main characteristics. The second, which we refer to as \textit{Goal-oriented Data Exploration} (GDE), is the process of exploring an already familiar dataset in light of a specific analytical goal or question, in order to derive specific, relevant insights. 

Numerous tools have been devised over the last decade for the purpose of assisting users in the manual, interactive process of data exploration~\cite{kraska2018northstar,googlesheet,li2021putting,deutch2020explained,tang2017extracting,eirinaki2014querie}. Most prominently, query recommendation systems and simplified analysis interfaces like Tableau~\cite{tableau} and Power BI~\cite{powerbi}. Recently, a new line of work called Automated Data Exploration (ADE), considers data exploration as a multi-step, AI \textit{control problem}~\cite{chanson2020traveling,chanson2022automatic,atenademo,atena_short,personnaz2021balancing,personnaz2021dora}. Unlike interactive tools that assist users step-by-step, Automated Data Exploration (ADE) systems take an input dataset and automatically generate a complete session of multiple, interconnected queries. Each query in the session builds on the results of one of the previous queries. The final output session is often displayed in a scientific notebook interface~\cite{perkel2018jupyter}, allowing users to quickly gain substantial knowledge on the data before investigating it further. 

Importantly, while existing ADE systems have been proven useful in assisting users in examining and familiarizing themselves with a new dataset, they are ineffective for the process of GDE. This is because existing ADE systems solve a predefined optimization problem, with a fixed objective function, thus always generating the same, or similar session for a given dataset. In the case of GDE, users need to answer a specific question, and seek insights that are relevant to their analytical goal. 
For illustration, consider the following example:

\begin{figure*}[t]
\vspace{-2mm}
    \includegraphics[width=\linewidth]{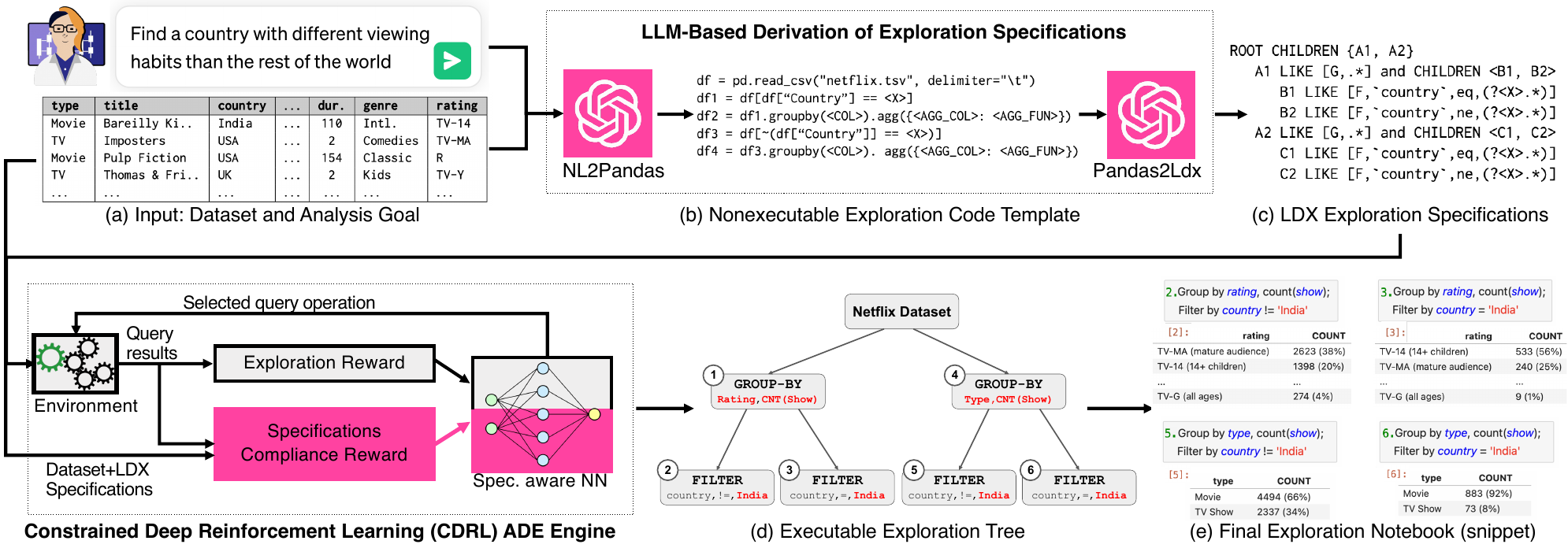}
    \vspace{-4mm}
    \caption{An Example \system{}\includegraphics[width=0.015\textwidth]{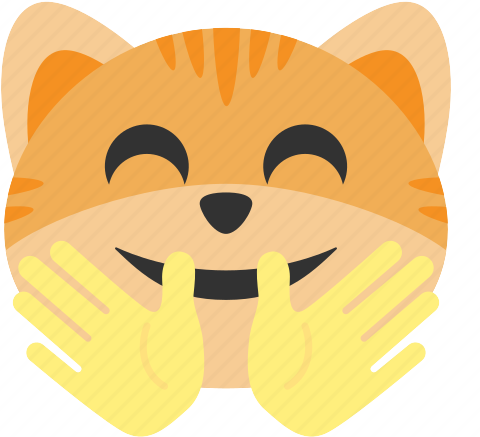} Workflow for Auto-Generating Goal-Oriented Exploration Sessions} 



    \label{fig:ap_workflow}
    \vspace{-2mm}
\end{figure*}

\begin{example}
\label{ex:intro}
Data Scientist Clarice, working at a media company, is assigned to analyze the Netflix Movies and TV Shows dataset~\cite{netflixkaggle}, which contains information about more than 8.8K different titles. Her current assignment is \textit{finding a country with atypical viewing habits, compared to the rest of the world} (to discover new insights that can be utilized to broaden the company's viewership audience). 
While Clarice is familiar with this dataset, she is tasked with a challenging analytical goal that cannot be answered via a single  query. To meet the goal, Clarice needs to examine countries in a trial-and-error manner, comparing them to the rest of the world with different attributes and aggregation functions. 

Using the existing ADE system ~\cite{atena_short}, Clarice receives an output exploration notebook, containing query results that imply generic insights such as ``\textit{Most Netflix titles originated in the US}''. However, these offer no help in respect to Clarice's analytical goal -  a specific question about countries with atypical viewing habits. 
\end{example}

To this end, we introduce \system{}, a \textit{Language-driven generative system for goal-oriented exploration}. \system{} is a novel ADE system that receives as input not only the user's dataset, but additionally, a description of the user's  \textit{analytical goal} in natural language. \system{} then generates a \textit{personalized}, exploratory session, containing queries that are tailored specifically to the dataset and the given goal at hand. 
\system{} follows two steps: First, it uses an LLM-based solution to interpret the input analysis goal and derives from it a set of specifications for the desired output exploration session. Second, the dataset along with the specifications are transferred to a novel, modular ADE engine which can adapt its output accordingly.

\begin{example}
\label{ex:intro2}
As depicted in Figure~\ref{fig:ap_workflow}, Clarice uploads the Netflix dataset to \system{} and types a description of her goal: \textit{``Find a country with different viewing habits than the rest of the world''}. \system{} then decides that the output exploration session should contain two comparisons of the same group-and-aggregate queries, one when filtering in on a country, and the second when that country is filtered out. These specifications are then inserted into \system{}'s modular ADE engine. The engine executes a multitude of exploration sessions until converging to an optimal one: Two group-by operations comparing the \textit{rating} and \textit{show-type} (using a \textit{count} aggregation), where each is employed on the results of two filter queries -- one by \texttt{Country=India}, and the second by \texttt{Country!=India}. Observing the output session notebook (See Fig.~\ref{fig:ap_workflow}e for a snippet) she quickly derives insights that are relevant to her goal, illustrating how India differs from the rest of the world: (1) \textit{``While the majority of titles in the rest of the world are rated \textit{TV-MA} (17+), in India, most titles are rated \textit{TV-14} (14+)''} and (2) \textit{``In India, the majority of titles are movies (93\%), whereas in the rest of the world, movies comprise only 66\% of the titles (with the rest being TV shows)''}.

\end{example}

\system{} is able to generate such goal-oriented sessions using two main components: A modular ADE framework that takes into consideration custom specifications, and an LLM-based solution for deriving such specifications from a natural language prompt. 



\vspace{1mm}
\noindent\textbf{1. Modular ADE framework with a dedicated specification language.}  Building an ADE framework for goal-oriented exploration requires two significant components lacking in existing ADE systems. First, a means to articulate custom exploration specifications, and second, the ability to integrate these specifications in the ADE optimization process. To this end, we first introduce \lang{}, a formal, intermediate language for data exploration. \lang{} allows to define the space of \textit{desired, relevant} exploration sessions with useful constructs for setting the structure, syntax, and the contextual relations between the query operations. Importantly, we devise an efficient \textit{verification} engine for \lang{}, which quickly determines if an output session is compliant with the specifications or not. 

Second, we develop a modular ADE engine that takes into account the input specifications when generating an output exploration session. We base our framework on ATNEA~\cite{atena_short}, an existing, goal-agnostic ADE system using Deep Reinforcement Learning (DRL). Our modular ADE engine contains two components necessary to support custom specifications: (1) A graduate \textit{\lang-compliance} reward signal, based on multiple variations of the \lang{} verification engine, used for providing a fine-grained numeric score which increases as the session is closer to satisfying all specifications. (2) A \textit{specification-aware} neural network architecture that derives its final structure from the \lang{} specifications. Our adaptive architecture is inspired by Constrained Deep Reinforcement Learning (CDRL) solutions~\cite{saunders2017trial,dalal2018safe} where the neural network agent is specifically designed to handle additional requirements, such as safety constraints in autonomous driving frameworks~\cite{garcia2015comprehensive}. In such systems, an external mechanism is used to override the agent's actions if they are violating the constraints. In our case, rather than overriding actions externally, we encourage the agent to perform compliant queries by dynamically shifting the action distribution probabilities toward queries that are more likely to be included in a specifications-compliant exploration session.

\vspace{1mm}
\noindent\textbf{2. LLM-Based solution for deriving exploration specifications from an analytical goal}. As previously mentioned, \system{} users specify their goal in natural language, meaning that they do not need to compose \lang{} specifications, but rather, these are derived directly from the analysis goal description. Our solution receives as input the user's goal as well as a short description of the dataset, and derives from it a syntactically correct \lang{} specification (This part is crucial as our modular ADE engine utilizes the \lang{} verification engine). Unlike more common tasks such as Text-to-SQL, for which LLMs demonstrate superior performance, for our task there is hardly any available resources in the LLMs' training data (see discussion in Section~\ref{sec:related}).
To overcome the absence of NL-\lang{} information in the LLM training data, we use a \emph{few-shot} setting~\cite{Wei2022ChainOT,nan2023enhancing}, coupled with \textit{intermediate code representation}~\cite{wolfson-etal-2020-break,Madaan2022LanguageMO,Chen2022ProgramOT}: Instead of directly instructing the LLM to generate \lang{} specifications, we adopt a two-stage prompting approach. In the first prompt, the LLM is tasked with expressing the specifications as a non-executable Python Pandas~\cite{pandas} code. In the second stage, an additional prompt instructs the LLM to translate the resulting code into formal \lang{} specifications. As LLMs are trained  on vast amounts of Python code, this intermediary step significantly improves their final performance.

\vspace{2mm}
\noindent\textbf{Experimental Evaluation \& Benchmark Dataset.} 
To evaluate \system{}, we constructed the first benchmark dataset, to our knowledge, for goal-oriented data exploration. Our benchmark contains 182 pairs of analytical goals and corresponding exploratory specifications, over three different datasets. We then conducted a thorough user study involving 30 participants to evaluate the relevance and overall quality of \system{} exploration sessions. We compared \system{} sessions to ones generated by ATENA~\cite{atena_short}, to sessions generated directly by ChatGPT~\cite{gpt35}, as well as to ones generated by the Google Sheets ML-Exploration tool~\cite{googlesheet}.
The results are highly positive: Sessions generated by \system{} were considered 1.5-2X more useful and allowed users to derive 3-5X more goal-relevant insights than the other automatic baselines.

\vspace{1mm}
A recent demo paper~\cite{lipman2023atena} briefly introduces \lang{} and an earlier prototype of its engine, with a main focus on a web interface for manual specification composition. In our current paper we present an end-to-end, tested solution that only requires the user to describe their analytical goal in natural language. 


\vspace{2mm}
\noindent\textit{Paper Outline.} We begin by surveying related work (§\ref{sec:related}), then formally define our problem and provide an example workflow of \system{} (§\ref{sec:prelim}). Next, we describe the \lang{} language (§\ref{sec:language}), our CDRL-based modular ADE framework (§\ref{sec:engine}) and the LLM-based solution for specifications derivation (§\ref{sec:ldx_llm}). Finally, we present our experimental evaluation (§\ref{sec:experiments}) and provide concluding remarks (§\ref{sec:conclusion}).



   \section{Related Work}
\label{sec:related}

\vspace{1mm}
\noindent\textbf{Assistive Tools for Interactive Exploration (Single Step).} 
Assisting users in data exploration has been the focus of numerous previous works.
Examples include simplified analysis interfaces for non-programmers~\cite{kraska2018northstar,googlesheet}, 
explanation systems for exploratory steps,~\cite{li2021putting,deutch2020explained}, insights extraction~\cite{tang2017extracting}, and recommender systems for \textit{single} exploratory steps~\cite{eirinaki2014querie,reactkdd,yan2020auto,joglekar2014smart,drosou2013ymaldb_full,sarawagi1998discovery,dimitriadou2016aide,di2019factorized}.
While these works facilitate the \textit{interactive} aspects of data exploration, \system{} focuses on a complementary dimension -- generating full exploratory sessions, joining the more recent line of research on ADE, as described below.

\noindent\textbf{Automated Data Exploration (ADE)}.
Rather than assisting users in formulating a single query, more recent systems such as~\cite{chanson2020traveling,chanson2022automatic,atenademo,personnaz2021dora,personnaz2021balancing} aim to generate an end-to-end exploratory process, given an input dataset, with the purpose of highlighting interesting aspects of the data, and providing thorough preliminary  insights. When presented in a notebook interface, such exemplar exploration sessions are highly useful for analysts and data scientists~\cite{rule2018exploration,kery2018story,perkel2018jupyter}.

 Due to the vast domain of possible exploration \textit{sessions}, systems such as~\cite{atenademo,atena_short,personnaz2021dora,personnaz2021balancing} use powerful optimization tools, such as dedicated \textit{deep reinforcement learning} (DRL) architectures and mathematical solvers. 
 However, as previously mentioned, these systems are \textit{agnostic} to the user's goal, due to their predefined objective function, which makes them generate the same session per dataset. \system{} is the first ADE framework, to our knowledge, designed for \textit{goal-oriented} exploration. 
 

\noindent\textbf{Data Visualization Specifications and Recommendations.} 
An adjacent research field focuses on assisting users in choosing appropriate data visualizations~\cite{vartak2015s,Zenvisage,lee2021lux,wongsuphasawat2016voyager}. While this is a crucial aspect of data exploration, it is complementary to our work which is focused on \textit{slicing and dicing} the data using filter and group-by queries. Systems such as~\cite{lee2021lux} can be used on the output sessions of \system{} to create compelling visualizations for each query result.

In addition, numerous specifications languages exist for data visualizations~\cite{hunter2007matplotlib,wongsuphasawat2017voyager,satyanarayan2016vega}.  
Similarly to these languages, the \lang{} language introduced in this work also uses parametric definitions for query operations. However, whereas visualization languages are used for defining a \textit{single visualization}, \lang{} focuses on defining a \textit{sequence} of exploration queries. 

\noindent\textbf{Text-to-SQL.}
As previously mentioned, \system{} uses an LLM-based solution that derives exploratory specifications from a textual description of the user's analytical goal. 
This task draws some similarities with 
the well-studied task of \textit{text-to-SQL}, where a structured query is translated from a natural language (NL) request~\cite{Affolter2019ACS,Kim2020NaturalLT}.
Recently, Text-to-SQL via LLMs~\cite{li2023llm,pourreza2023dinsql} has shown promising results, nearly comparable to dedicated architectures~\cite{Zhong2017Seq2SQLGS}.
This is mainly due to the existing resources used for this task, such as supervised datasets like \cite{Zhong2017Seq2SQLGS,yu2018spider,li2023llm}, a plethora of academic papers and books, as well as practical tips in programming internet forums. Differently, solving our new task of NL-to-\lang{} requires overcoming additional challenges: (1) LLMs are not explicitly trained on vast amounts of exploratory sessions, (2) Our task requires specifying a sequence of interconnected queries, rather than a single SQL query, and therefore more difficult to derive solely based on a description of the task and dataset, and (3) rather than generating the full session, the LLM is tasked with \textit{partially} specifying it, thus leaving some of the query parameters to be discovered by the ADE engine. In Text-to-SQL, the NL request is instantly translated to an executable query. We show in Section~\ref{sec:experiments} that exploratory sessions generated \textit{directly} by ChatGPT are significantly inferior to sessions generated by \system{}.


\vspace{1mm}
\noindent\textbf{LLM Applications in Data Management.}
The promising generative results obtained by LLMs paved the path to recent research works investigating how to utilize it for data management tasks. Applications (beyond text-to-SQL) include data integration \cite{Arora2023LanguageME}, table discovery \cite{Dong2022DeepJoinJT}, columns annotation~\cite{suhara2022annotating} and even the potential of substituting database query execution engines \cite{Thorne2021FromNL,Narayan2022CanFM,Saeed2023QueryingLL,Trummer2023DemonstratingGG}. These exciting research directions are orthogonal to our work.  

    \section{Problem Setting, Example Workflow}
\label{sec:prelim}
We first define the problem of goal-oriented, automated data exploration, then present an example workflow of \system{}.

\paragraph*{The goal-oriented ADE problem}
Given an input dataset $D$, and an analytical goal description $g$,
we define the task of automatically generating a full exploration session comprised of $N$ queries: $q_1,q_2,\dots q_N$.  As in standard ADE settings, we assume a predefined set of query operation types. Following~\cite{atena_short} we focus on the following \textit{parametric} filter, and group-by query operations: A \textit{filter} operation is defined by \verb![F,attr,op,term]!, where \verb!attr! is an attribute in the dataset, \verb!op! is a comparison operator  (e.g. $=, \geq, contains$) and \verb!term! is a numerical/textual term to filter by. A \textit{Group-and Aggregate} operation is defined by \verb![G,g_attr,agg_func,agg_attr]!, i.e., grouping on attribute \verb!g_attr!, aggregating on attribute \verb!agg_attr! using an aggregation function \verb!agg_func! (we discuss the support of additional operations below).

We further assume a \textit{tree-based exploration} model, following~\cite{atena_short,reactkdd}, in which
each query operation $q_i$ is represented by a node, and is applied on the results of its \textit{parent} operation.
The ``root'' node of the exploration tree is the original dataset before any operation is applied,
and the query execution order corresponds with the tree pre-order traversal (See Figure~\ref{fig:ap_workflow}d for an example exploration tree). For dataset $D$, we denote an exploration tree by $T_D$.

Now, given a \textit{utility score function} for an exploratory session, denoted $U$,
a goal-agnostic ADE system is tasked to generate a session $T_{D}$ such that $U(T_{D})$ is maximal. Multiple such notions are defined in previous work~\cite{atena_short,personnaz2021dora,chanson2020traveling}.
In \system{}, given a dataset $D$ and the goal $g$, our objective is to generate a session 
$T_{D}$ such that $U(T_{D})$ is maximal \textit{and} relevant, w.r.t. analysis goal $g$. In \system{}, the relevance of a session is determined according to a set of exploration specifications $\lquery$, derived w.r.t. the goal $g$. If $\lquery(T_{D}) = True$, i.e., the session is compliant with the specifications, then we say it is \textit{relevant} w.r.t. goal $g$.



\label{sec:sysoverview}

\paragraph{Example workflow}
Figure~\ref{fig:ap_workflow} illustrates the detailed architecture and an example workflow of \system{}, extending Example~\ref{ex:intro2}. The user uploads a dataset and an analytical task (Fig.~\ref{fig:ap_workflow}a), then \system{} works using a two step process: (1) It first derives a set of exploration specifications $\lquery$ that form a ``skeleton'' for the output session. This skeleton accommodates a variety of compatible instances. In the second step (2), our \textit{Modular ADE engine} generates the full session $T_D$, which maximizes the exploration score $U(T_{D})$ (we use the notion from~\cite{atena_short}, as explained below) and is also compliant with the derived specifications $\lquery$.  


\vspace{1mm}
\noindent\textbf{Step 1: Deriving Exploration specifications w.r.t. the goal.}
We use an LLM-based solution to derive exploration specifications from $D$ and $g$, expressed in \lang{} (described in Section~\ref{sec:language}).

As mentioned above, we use a two-stage prompting approach, to overcome the absence of relevant explicit knowledge in the LLM training data. First, we prompt the LLM to generate \textit{non-executable Python Code}, as
depicted in Figure~\ref{fig:ap_workflow}b. Note that this is merely an intermediate gateway for expressing the specifications, as this code cannot be executed. In particular, it contains special placeholders (marked with $<>$) for query parameters that will be later instantiated by the ADE engine, in a manner that maximizes the general exploration score. As described in Example~\ref{ex:intro2}, \system{} takes the goal of finding an atypical country in the Netflix dataset, and derives that the output session should contain filter operations on the `Country' column -- one for a specific country, and the other for the complement data (i.e., the rest of the world), each followed by the same group-and-aggregate operation. See that the specific country and the group-by parameters are \textit{not} specified. These will be instantiated later by the modular ADE engine, which will discover the instances that maximize the exploration utility. 
Last, the non-executable code is then translated to \lang{} via a subsequent prompt, as illustrated in Fig.~\ref{fig:ap_workflow}c. Returning syntactically correct \lang{} is crucial, as the \lang{} verification engine (Section~\ref{sec:compliance_verification}) is embedded in the ADE optimization process, as explained in Section~\ref{sec:engine}. 


\vspace{1mm}
\noindent\textbf{Step 2: Generating a maximal-utility exploratory session, in accordance with the goal-driven specifications.} In the second step our modular ADE framework performs a CDRL process, as illustrated in Figure~\ref{fig:ap_workflow} (bottom left): optimizing a generic exploration reward (defined in~\cite{atena_short})  while ensuring that the output session is compliant with the input specifications. This is enabled due to our \textit{compliance reward scheme} (Section~\ref{sec:reward}) that employs multiple instances of the \lang{} verification engine, and a novel \textit{specification-aware neural network} which adjusts its structure based on the input specifications (Section~\ref{sec:network}).

After the CDRL process converges, \system{} produces an \textit{executable exploration tree} (Fig.\ref{fig:ap_workflow}d), consisting of executable query operations that adhere to the input specifications while maximizing the generic exploration score.  
The query parameters marked in red are the ones discovered by the CDRL engine: the country filter value <X> is \textit{`India'}, and the comparison involves a \textit{count} aggregation over the attributes \textit{rating} and \textit{show type}. This exploratory session is then presented to the user as a \textit{scientific notebook}, as depicted in Fig.\ref{fig:ap_workflow}e. The notebook snippet demonstrates that the output exploratory session indeed reveals interesting and relevant insights, illustrating how India differs from the rest of the world in terms of the title \textit{ratings} and \textit{types}, as explained in Example~\ref{ex:intro2}. 

\paragraph*{Limitations and Scope}
\vspace{2mm}
We conclude with three remarks on the scope and limitations of \system{}: 

\vspace{1mm}
\noindent\textit{1. ADE Vs. interactive exploration.} 
Recall that \system{} is an ADE system designed to generate comprehensive exploratory sessions, similar to the approaches in~\cite{chanson2020traveling,chanson2022automatic,personnaz2021dora,personnaz2021balancing}. As discussed in Section~\ref{sec:related}, ADE systems are not intended to replace interactive exploration tools~\cite{eirinaki2014querie,reactkdd,yan2020auto}. Instead, their primary role is to be used \textit{before} users engage in interactive data exploration, offering valuable, thorough preliminary insights. This preparatory step is akin to reviewing \textit{human-generated} exploration notebooks found on platforms like Kaggle and Github~\cite{kery2018story}, providing a solid foundation for subsequent analysis. 
Naturally, due to the vast search space, the output of ADE systems~\cite{atena_short,chanson2020traveling} may take several minutes to generate (see the discussion in Section~\ref{sec:exp_performance}). However, this longer running time is acceptable, given that ADE systems are not intended for real-time interaction but for providing thorough, preparatory insights.


\vspace{1mm}

\noindent\textit{2. Supported Query Operations.} \system{} currently supports filter and group-and-aggregate queries, which are fundamental for many exploratory data analysis tasks. While the system does not yet support join or union operations, this limitation is intentional to ensure robustness and efficiency in its core functions. Extending the framework to include these additional operations is feasible, requiring only minor modifications to the DRL environment: adding a parametric definition of the new operation and establishing a corresponding utility notion (e.g., for join, such notions are detailed in~\cite{deutch2022fedex,zhu2019josie}). We recognize the importance of these operations and plan to integrate them in future work.


\vspace{1mm}

\noindent\textit{3. Future Extension: Spelled-out Insights and Visualizations.} 
Within its current scope, \system{} generates a sequence of query operations that align with the user's analytical goals (see an example output in Figure~\ref{fig:ap_workflow}e). While the generated queries are designed to be intuitive and easy to interpret (as evidenced by our user study results in Section~\ref{sec:exp_us}), we acknowledge that some users may prefer more compelling outputs, such as natural-language insights or visualizations. This is an important area for future enhancement. We plan to integrate auto-visualization systems like LUX~\cite{lee2021lux} and explore the use of LLMs to provide natural language summaries of the exploratory sessions. 

    \section{Exploration Specification Language}
\label{sec:language}



We first describe \lang{}, our intermediate language designed for explicitly defining a \textit{sub-space} of exploration sessions that can be relevant for the input analysis goal. Importantly, we further introduce an efficient verification algorithm for \lang{}, which is embedded in our ADE engine (Section~\ref{sec:reward}).  






\subsection{\lang{} Language Overview}
\label{sec:lang_overview}
Recall that an exploration session tree $T_D$ comprises a sequence of query operations, where each query $q_i$ is employed on the results of one of the previous queries $q_j, j<i$. 
\lang{} therefore allows posing specifications for (1) the session structure, i.e., the shape of the tree which reflects the execution order and the input data for each query, (2) the parameters and type of the queries, and (3) \textit{continuity variables} for controlling how queries are interconnected. The latter aspect is particularly important for exploration sessions examined by users, as the semantic connection between the queries allows building  an exploration\textit{ narrative}~\cite{kery2018story,perkel2018jupyter} that gradually leads the viewer to nontrivial insights on the data.




Our specification language \lang{} extends Tregex~\cite{levy2006tregex}, a query language for tree-structured data\footnote{Tregex natively allows partially specifying structural properties of the tree, as well as the nodes' labels, yet is missing the definitions of \textit{continuity}.}. 
The basic unit in \lang{} is a \textit{single node specification}, which addresses a particular node (query operation, in our context). A full \lang{} specifications query is then composed by conjuncting multiple single-node specifications, interconnected using the \textit{continuity variables}. 
We begin with a simple ``hello world'' example, then describe \lang{} constructs in more detail.

\begin{example}
\label{ex:lang}
The following \lang{} query describes a simple exploration session ``skeleton'' with two query operations: a group-by, followed by a filter operation, both employed on the full dataset (the root node in $T$). It also specifies that the filter is to be performed on the same attribute as the group-by. The rest of the parameters are left \textit{unspecified}.

\begin{center}
\begin{BVerbatim}
ROOT CHILDREN <A,B>
    A LIKE [G,(?<X>.*),.*]
    B LIKE [F,(?<X>.*),.*]
\end{BVerbatim}
\end{center}

The query contains three \textit{named-nodes} -- \e{ROOT}, \e{A} and \e{B}, each is differently specified. The \e{ROOT} node represents the raw dataset, has two immediate children \e{A} and \e{B} -- the  group-by and filter operations (both use it as input data). \e{A} is a group-by with ``free'' parameters: unspecified group attribute, aggregation function, and aggregation attribute, and \e{B} is a filter operation  with unspecified operator and term (Recall the parametric definition of queries in Section~\ref{sec:prelim}). Unspecified parameters are marked with \e{*}, but see that the \textit{attribute} parameter in both query operations is marked with \e{(?<X>.*)}. This means that \e{X} is a \textit{continuity variable} that ensures that both operations need to use \textit{the same} column parameter.
\end{example}


We next briefly describe the constructs of \lang{} (Full description and more examples are provided in~\cite{our_github_repository}).

\vspace{1mm}
\noindent\textbf{Specifying exploration tree structure.} 
The session structure is specified via tree-structure primitives such as \verb!CHILDREN! and \verb!DESCENDANTS!.
For instance, \verb!'A CHILDREN <B,+>'! states that Operation \verb!A! has a subsequent operation named \verb!B!, and at least one more (unnamed) operation, as indicated by the \e{+} sign.
Importantly, recall that the fact that \e{B} is a child of \e{A} not only means that Operation \e{B} was executed after Operation \e{A}, but also that \e{B} is employed on the results of Operation \e{A} (i.e., rather than on the original dataset). 

\vspace{1mm}
\noindent\textbf{Specifying query operation parameters.} \lang{} allows for partially specifying the operations using \textit{regular expressions} (regex), as they define match patterns that can cover multiple instances. For example, the expression \verb!'A LIKE [G,'country',SUM|AVG, *]'! specifies that Operation \e{A} is a \textit{group-by} on the attribute \textit{country}, showing either \textit{sum} or \textit{average} of \textit{some} attribute (marked with $*$). 

\vspace{1mm}
\noindent\textbf{Continuity Variables.}
We next introduce the continuity variables in \lang{}, which allow constructing more complex specifications that \textit{contextually} connect between operations' free parameters once instantiated.
\lang{} allows this using named-groups~\cite{aho1991algorithms} syntax. Yet differently than standard regular expressions, which only allow ``capturing'' a specific part of the string, in \lang{} these variables are used to constrain the operations in subsequent nodes.
For instance, the statement `\verb!B1 LIKE [F,'country',eq,.*]!' (taken from the \lang{} query in Figure~\ref{fig:ap_workflow}c) specifies that Operation \e{B1} is an \textit{equality filter on the attribute \textit{`country'}, where the filter term is free}. To capture the filter term in a continuity variable we use 
named-groups syntax: `\verb!B1 LIKE [F,'country',eq,(?<X>.*)]!' --
in which the free filter term (\texttt{.*}) is captured into the variable \texttt{X}.
Using this variable in subsequent operation specifications will restrict them to the same filter term (even though the term is not explicitly specified). For instance, as shown in Figure~\ref{fig:ap_workflow}c, the subsequent specification is `\verb!B2 LIKE [F,'country',neq,(?<X>.*)]!', indicating that the next filter should focus on all countries \textit{other} than the one specified in the previous operation.


\subsection{\lang{} Verification Engine}
\label{sec:compliance_verification}
We next describe our \lang{} verification engine, which takes an exploration session tree $\stree$ and a \lang{} specifications query $\lquery$, and verifies whether $\stree$ is compliant with $\lquery$.

For an input \lang{} query $\lquery$, we denote the set of its \textit{named nodes} (e.g., nodes \e{ROOT}, \e{A} and \e{B} in Example~\ref{ex:lang}) by $Nodes(\lquery)$, and the set of its continuity variables by $\conts$. 
We first define an \lang{} assignment,
then describe our verification procedure that searches for valid assignments.

\begin{definition}[\lang{} Assignment]
\label{def:assignment}
Given an \lang{} query $\lquery$ and an exploration session tree $\stree$, an \textit{assignment} $A(\lquery,\stree)= \langle
\phi_V, \phi_C \rangle$, s.t.,
(1) $\phi_V$ is a \textit{node mapping function}, assigning each named node $n \in Nodes(\lquery)$ an operation node $v \in V(\stree)$ in the exploratory session $T$. (2) $\phi_C$ is a \textit{continuity mapping function}, assigning each continuity variable $c \in Cont(\lquery)$ a possible value. The initial node mapping is $\phi_V(\e{ROOT})=0$, i.e., mapping the root node in the \lang{} query to the root node of  $\stree$.

\end{definition}




\paragraph*{\lang{} Verification Algorithm}
Recall that an \lang{} query $\lquery$ comprises a set of \textit{single node specifications},
s.t. each specification $\spec \in \slist$ refers to a single named node in $\lquery$.  We denote the named node of $\spec$ by $Node(s)$, and the (possibly empty) set of continuity variables in $\spec$ by $Cont(\spec)$. 
The \lang{} verification algorithm, as depicted in Algorithm~\ref{alg:verification}, 
takes as input an \lang{} specifications query $\slist$, an exploration tree $\stree$, and the initial assignment $A$,
in which $\phi_V$ contains the initial root mapping (Definition~\ref{def:assignment}) and an empty continuity mapping $\phi_C$,
and returns \e{true} if there exists at least one valid assignment $A(\lquery,\stree)$.  
Note that since Tregex does not support continuity variables, we can only use its node matching function  \e{GetTregexNodeMatch}~\cite{tregeximplementation} in our algorithm. This function, as described in~\cite{tregeximplementation}, 
takes as input a single specification $s$, a tree $T$, and the current node mapping $\phi_V$ and returns all valid node matches for $Node(s)$, denoted $V_T^s$, given the current state of the node mapping $\phi_V$. 

\vspace{1mm}
Our verification procedure, as described in Algorithm~\ref{alg:verification} works as follows. In each recursive call, a single specification $s$ is popped from $\slist$ (Line~\ref{ln:pop}). Then, $s$ is updated with the continuity values according to $\phi_C$ (Lines~\ref{ln:for_cont1}-\ref{ln:update_s}): if a continuity variable $c$ is already assigned a value in $\phi_C$, we update the instance of $c$ in $s$, denoted $s.c$, with the corresponding value $\phi_C(c)$.
Next (Line~\ref{ln:get_tregex}), when all available continuity variables are updated in $s$, we use the Tregex \e{GetTregexNodeMatch} function, to obtain a set $V_T^s$ of possible valid assignments for $Node(s)$. 
Then, for each node $v \in V_T^s$, 
we first update the node mapping $\phi_V$ (Line~\ref{ln:update_phi_v}) and the continuity mapping $\phi_C$ (Lines~\ref{ln:for_cont2}-\ref{ln:cont_Assignment}): we assign each continuity variable $c$ the concrete value of $c$ from $v$, denoted $v.c$. (Recall that $v$ already satisfies $s$ also w.r.t. $\phi_C$, therefore only unassigned variables in $Cont(s)$ are updated.) 
Once both mappings are updated (denoted $\phi_V^s$ and $\phi_C^s$), we make a recursive call to \textit{Verify\lang{}}
(Line~\ref{ln:recursive}), now with the shorter $\slist$ (after popping out $s$) and the new mappings ($\phi_V^s$, $\phi_C^s$).
Finally, the recursion stops in case there is no valid assignment (Line~\ref{ln:false}) or when  $\slist$ is finally empty (Line~\ref{ln:true}). 
In Section \ref{sec:engine} we describe how multiple variations of the \lang{} verification algorithm are used within the optimization process of our modular ADE engine.  


    \section{CDRL Framework for Modular ADE}
\label{sec:engine}
Recall that ADE systems optimize over the domain of all possible exploration sessions, thus requiring powerful optimization tools~\cite{chanson2022automatic,atena_short,personnaz2021balancing}. We base our modular ADE engine on the goal-agnostic Deep Reinforcement Learning (DRL) framework for data exploration presented in~\cite{atena_short}. In the DRL setting, a neural-network agent produces a maximal-scoring session (using a predefined exploration reward function) by employing a multitude of intermediate sessions, then updating its internal policy according to their obtained scores until converging to an optimal one.

Different than~\cite{atena_short}, our \textit{modular} ADE framework takes a given dataset $D$, as well as \lang{} specifications $\lquery$, and generates a high-scoring exploration session $\stree$ which is in compliance with $\lquery$. 
The main challenge which arises here is to effectively embed the specifications as a part of the optimization process.  A naive integration would have been to incorporate, in addition to the generic exploration score, a binary score derived from the result of the verification engine for each generated session (i.e., compliant/non-compliant). However, this naive solution introduces a \textit{reward sparsity} problem \cite{mataric1994reward}, a prominent challenge in reinforcement learning arising when the agent scarcely obtains a positive feedback, thus failing to converge. Our experimental evaluation in Section~\ref{sec:exp_performance} indeed shows that such a solution fails to converge on \textit{all} tested \lang{} queries. We next overview our solution, based on Constrained Deep Reinforcement Learning (CDRL)~\cite{saunders2017trial,dalal2018safe}.

\paragraph{CDRL Framework Overview}
To effectively embed the specifications in the optimization process we use a twofold solution:
 First, we introduce a flexible compliance reward scheme that gradually guides the DRL agent towards fully compliant sessions by encouraging it to first generate structurally compliant sessions (learning the queries type and order of execution), and only then refine individual query parameters. Then, we devise a novel neural network architecture, inspired by intervention-based CDRL~\cite{saunders2017trial,dalal2018safe}. In these CDRL systems an external mechanism is used to override the agent's actions if they are violating the constraints. In our case, we cannot always detect a violation immediately, and verify the compliance only at the end of a session. Thus, 
 rather than overriding actions externally, we internally encourage the agent to perform compliant operations via a novel \textit{specification-aware} network architecture, pushing query parameters that are likely to comply with the specifications with a higher probability. We show in Section~\ref{sec:exp_performance} that only the combination of these two solutions allow \system{} to successfully and consistently converge.

\SetCommentSty{rmfamily}
\setlength{\textfloatsep}{2pt}
 \IncMargin{1em}
\begin{algorithm}[t]
\small 
\DontPrintSemicolon
   \SetNlSkip{2mm}
   \SetKwProg{myfunc}{Verify\lang}{
   }{}
   \nonl \myfunc{$(\stree,\slist, A= \langle \phi_V=\{\texttt{ROOT:0}\},\phi_C=\emptyset)\rangle$ \quad~\quad~\textup{\textrm{//~Inputs: Exploration tree $\stree$, \lang{} Specifications $\slist$, assignment $A$}}}{
   
    \lIf{$\slist=\emptyset$}{\Return True} \label{ln:true}
    $s \leftarrow \slist.pop()$ \tcp*{pop a single node specification from $\slist$} \label{ln:pop}
    
    \label{ln:for_cont1}  \For(\tcp*[f]{Assign continuity vars in $s$}){$c \in Cont(s)$}{ 
       \lIf{$c \in \phi_C$}{$s.c \leftarrow \phi_C(c)$} \label{ln:update_s}  
    }

    $V_T^s \leftarrow \textbf{GetTregexNodeMatches}(s,T,\phi_V)$ \label{ln:get_tregex}
    
   \label{ln:forverify}  \For{$v \in \mathcal{V}_T^s$ }{
   $\phi_V^s \leftarrow \phi_V \cup \{Node(s):v\}$, $\phi_C^s \leftarrow \phi_C$ \label{ln:update_phi_v}
   
    \label{ln:for_cont2}  \For(\tcp*[f]{Update continuity mapping}){$c \in Cont(s)$}{ 
       $\phi_C^s(c) \leftarrow v.c$
        \label{ln:cont_Assignment}  
    }

       \If{$\textbf{Verify\lang}(\stree,\slist, \langle \phi_V^s, \phi_C^s \rangle$) \label{ln:recursive} }{\Return True}   
    }

 \Return False \label{ln:false}}{} 
    

\caption{\lang{} Query Compliance Verification}
 \label{alg:verification}
\end{algorithm}

We next define the Markov Decision Process (MDP) model used in our CDRL framework, then delve into the \lang{}-compliance reward scheme and specification-aware network.  

\subsection{MDP Model}
Following~\cite{atena_short} we use an episodic MDP model in our CDRL engine, defined as 
$\mathcal{M} \coloneqq \left( \mathcal{S}, 
\mathcal{A}, 
\Delta_a, 
R_a \right)$, where $\mathcal{S}$ is a state space; $\mathcal{A}$ is an action space;
$\Delta_a:\mathcal{S} \times \mathcal{A}\rightarrow \mathcal{S}$ is a transition function that returns the outcome state $S'$ obtained from employing an action $a$ in state $S$; and $R_a(S,a)$ is the \textit{reward} received for action $a$ in state $S$.

Our MDP model is defined as follows: Given a dataset $D$ and specifications $\lquery$, the agent produces an exploration session $\stree$ on $D$ in each episode. At each step $i$, the agent employs a parametric query operation $q_i$, as defined in Section~\ref{sec:prelim}. After executing an operation, the agent transitions to state $S_i = \Delta_a(S_{i-1},a)$, where $S_i$ represents the resulting view of $q_i$.
In addition to query operations, the agent can use a \textit{back} operation to return to a previous state and start a new action from there. 

For each action, the agent receives a bi-objective reward: $$R_a(S_i,a) \coloneq \alpha \cdot R_{gen}(S_i,a) + \beta \cdot R_{comp}(S_i,a,\lquery)$$ 
The first reward component $R_{gen}$ is based on the generic exploration reward defined in~\cite{atena_short}. It is a weighted sum of the interestingness scores of the session's individual queries and their diversity: $ R_{gen}(S_i,a) = \mu \cdot \sum_{j \leq i} \text{Interestingness}(q_j) + \lambda \cdot \text{Diversity}(S_i)$. As described in~\cite{atena_short}, interestingness scores are calculated using \textit{KL-divergence} for filter operations and \textit{conciseness}~\cite{geng2006interestingness} for group-by-and-aggregate operations. The diversity of the session $S_i$ is measured by computing the minimal distance between $q_i$ and a previous query $q_j, j\leq i$ (using the query results distance provided in~\cite{atena_short}).

The second component, $R_{comp}(S_i,a,\lquery)$, is a compliance reward unique to \system{}. This reward is based on the input \lang{} specifications $\lquery$ and is described in more detail below.

\subsection{\lang{}-Compliance Reward Scheme}
\label{sec:reward}
Given \lang{} specifications $\lquery$, and an exploratory session $\stree$, we define our \textit{compliance} reward signal, received at each step $i$: 
$$R_{comp}(S_i,a,\lquery)\coloneq \gamma\cdot EOS(S_i,a,T,\lquery)+\delta \cdot IMM(S_i,a,\stree^i,\lquery)$$ 

Here $EOS$ is an \textit{end-of-session} feedback, equally divided across all query operations; and $IMM$ is received immediately, per operation. 
These signals provide a fine-grained feedback, allowing our CDRL engine to overcome the \textit{reward sparsity problem} described above. 

\paragraph{End-of-Session Compliance Reward} 
The EOS reward component $EOS(S_i,a,T,\lquery)$ is received at the end of an episode (once the exploration session $\stree$ is fully generated), then equally distributed across all states $S_i$.
  We utilize the \lang{} verification engine (Algorithm~\ref{alg:verification}), but in light of the observation that \textit{structural specifications should be learned first}. Intuitively, if the agent learns to generate \textit{correct} query operations in an \textit{incorrect} order/structure, the learning process becomes largely futile as reordering requires the agent to relearn the session from scratch. We therefore partition the set of individual node specifications in $\lquery$ to
  \textit{structural} and \textit{operational} subsets, denoted by $struct(\lquery)$ and $opr(\lquery)$, s.t. $struct(\lquery) \cup opr(\lquery) = \lquery$. $struct(\lquery)$ refer to the definitions of the session tree structure and $opr(\lquery)$ to the definition of query operation parameters, as described in Section~\ref{sec:language}. 

Briefly, our end-of-session rewards works as follows (See Appendix~\ref{app:reward} for full details). 
First, we use Algorithm~\ref{alg:verification} to check if $\stree$ complies with $\lquery$. Then, a conditional reward is granted, according to the following three cases: (1) If fully compliant, a high positive reward is given. (2) If $\stree$ is not compliant with $\lquery$, we check its compliance only with $struct(\lquery)$, the structural specifications in $\lquery$. If no valid assignments are found, a fixed negative penalty is applied. (3) If $\stree$ satisfies $struct(\lquery)$ but not $opr(\lquery)$, i.e., the operational specifications, a non-negative reward is assigned based on the number of satisfied query parameters (The larger the number of satisfied parameters, the higher the reward).
Intuitively, this reward enforces the learning of correct structure by imposing a high penalty for non-compliant sessions. Once the correct structure is learned, the agent receives gradually increasing rewards to encourage satisfaction of operational specifications. Upon generating a fully compliant session, the agent receives a high positive reward.




\begin{figure}[t]
\vspace{-4mm}
    \includegraphics[width=0.95\linewidth]{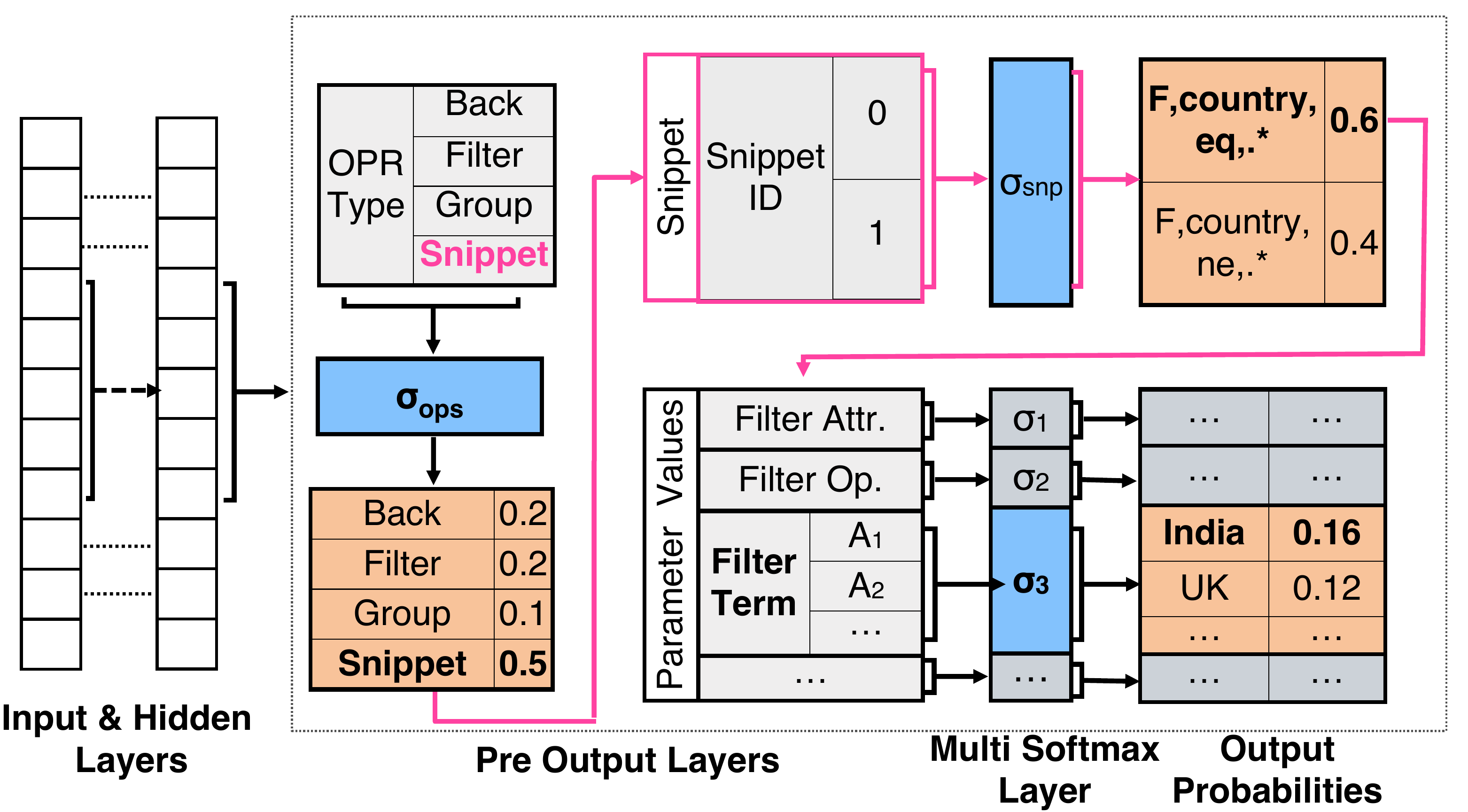}
    \vspace{-3mm}
    \caption{Specification-Aware Network Architecture}
    \label{fig:ButtonsArchitecture}
\end{figure}

\vspace{1mm}

\noindent\textit{Immediate (per-operation) Compliance Reward.} To reinforce adherence to structural 
constraints, we introduce an \textit{immediate} reward signal $IMM(S_i,a,\stree^i,\lquery)$ granted 
individually for each step $i$. This real-time signal negatively rewards specific operations that 
violate the structural specifications $struct(\lquery)$. 
To do so, we use a modification of the \lang{} verification engine (Algorithm~\ref{alg:verification}), that can operate on an \textit{ongoing} session $\stree^i$ (in step $i$) rather than a full session $\stree$. 
Intuitively, we assess the possibility of a \textit{future} assignment satisfying $struct(\lquery)$ in up to $N-1$ more steps. This is done by attempting to extend the exploration tree with $N-i$ additional ``blank'' nodes, respecting the order of query operations execution. In case no valid assignment is found to any of the new trees, a negative reward is granted. The number of possible tree completions throughout an $N$-size session is bounded by $C_N$, the Catalan number (See Appendix~\ref{app:reward} for full details). In practice, we show in Section~\ref{sec:exp_performance} that this reward poses a negligible computational overhead on the optimization process.

\subsection{Specification-Aware Neural Network}

\label{sec:network}
\label{sec:network_arch}
We next describe our specification-aware architecture used to increase the probability of choosing compliant operations. Our neural network modifies its structure according to the input \lang{} specifications, by creating special segments for operation ``snippets'' likely to be compatible with \lang{}. The agent can uses these snippets more frequently, thus advance faster toward a fully compliant session.

Figure~\ref{fig:ButtonsArchitecture} depicts the network architecture (Specifications-aware functionality is highlighted in pink). First, rhe input layer receives an observation of the current state $S_i$ in the MDP model, and passes it to the dense hidden layers. Then, the agent composes a query operation  via the pre-output layers, where it first chooses an operation type and subsequently its corresponding parameters. As depicted in Figure~\ref{fig:ButtonsArchitecture}, Softmax Segment $\sigma_{ops}$ is connected to the operation types, and Segments $\sigma_1,\sigma_2, \dots$ are connected to the value domain of each parameter. 

In our architecture, we add a new high-level action, called ``snippet'' ($\sigma_{snp}$).
When choosing this action, the agent is directed to select a particular snippet that is derived from the operational specifications $opr(\lquery)$. The snippets function as operation ``shortcuts'', which eliminate the need for composing full, compliant operations from scratch. For example, using a snippet of \e{`F, Country, eq`} (See Fig.~\ref{fig:ButtonsArchitecture}), only requires the agent to choose a filter term, rather than composing the full query operation.

Given an \lang{} qery $\lquery$, the network architecture is derived as follows. First, we generate an individual \textit{snippet} neuron for each operational specification $s \in opr(\lquery)$ (In case the regular expression in $s$ contains a disjunction, we generate an individual snippet for each option) 
All snippet neurons, as depicted in Figure~\ref{fig:ButtonsArchitecture}, are connected to $\sigma_{snp}$, the snippet multi-softmax segment. 
Now, to choose the ``free'' parameter, unspecified in $s$, the snippet neuron is wired to the corresponding parameters in the multi-softmax segments. For instance, the snippet of \e{`F, Country, eq`} is wired to $\sigma_3$ for choosing a filter \textit{term} parameter, as depicted in Fig.~\ref{fig:ButtonsArchitecture}. In Appendix~\ref{ex:network} we provide an example such derivation process.

In combination with the reward scheme presented earlier, our network architecture allows \system{} to consistently generate compliant exploration sessions in 100\% of the datasets and \lang{} queries in our experiments, as detailed in Section~\ref{sec:exp_performance}.

    \section{LLM-based solution for deriving exploration specifications}
\label{sec:ldx_llm}

As previously mentioned, \system{} users do not need to manually compose \lang{} specifications, but only provide a description of their analytical goal $g$. Deriving \lang{} specifications for the CDRL engine is done via an LLM-based solution: 
We use a \emph{few-shot} setting, renowned for its excellent performance across diverse analytical goals~\cite{Wei2022ChainOT,nan2023enhancing}. In this approach, several illustrative examples are provided to the LLM before soliciting task completion. Then, to further enhance the distillation of exploration specifications, we also use a solution based on \textit{intermediate code representation}~\cite{Gao2022PALPL,Madaan2022LanguageMO,Chen2022ProgramOT,Zhang2023ExploringTC}, where instead of directly instructing the LLM to generate \lang{} specifications, we adopt a two-stage chained prompt: In the first prompt, the LLM is tasked with expressing the specifications as a non-executable, template Python Pandas~\cite{pandas} code, restricted to the operations supported by \system{}. The template code (See Figure~\ref{fig:ap_workflow}a) contains special placeholders representing the query operations (or specific parameters) to be discovered in a data-driven manner. In the second stage, an additional prompt instructs the LLM to translate the intermediate Pandas code into formal \lang{} specifications. 
We coin our approach \textit{NL2PD2LDX}.

Recall again that the last conversion to \lang{} is \textit{required} in \system{}, due to its efficient verification engine embedded in the CDRL process.  
As we empirically show in Section~\ref{sec:llm_exp}, our two-stage approach exhibits superior generalization compared to a direct NL-to-\lang{} approach, and when combined with the CDRL engine described above, it produces exploratory sessions that are deemed more useful and insightful than other baselines such as ChatGPT and ATENA~\cite{atena_short}.

\paragraph*{Prompt Engineering}

\begin{figure}[t]
\vspace{-2mm}
    \includegraphics[width=0.98\linewidth]{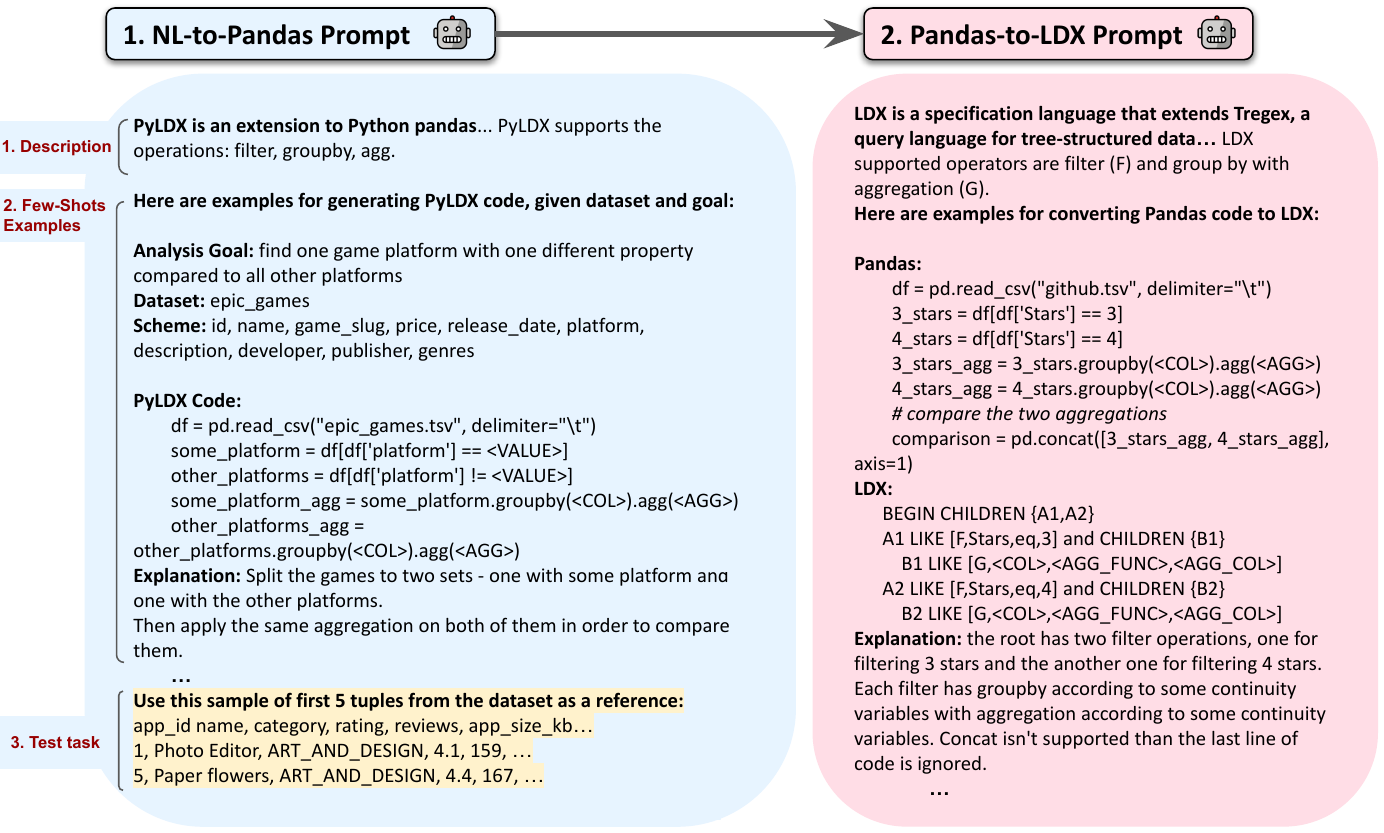}
    \vspace{-3mm}
    \caption{Examples of the chained prompts: (1) NL to non-executable Pandas code, and (2) Pandas code to \lang{}}
    \label{fig:prompt_pandas}
\end{figure}

Figure~\ref{fig:prompt_pandas} depicts a snippet of our chained prompts: NL-to-Pandas and Pandas-to-LDX.

\vspace{1mm}
\noindent\textit{NL-to-Pandas.} The prompt is structured into three main components: (1) NL-to-Pandas task description; (2) a series of few-shot examples; (3) the test analysis goal alongside a small dataset sample.
Each few-shot example in (2) comprises several steps: (a) example analytical goal; (b) dataset and schema description (e.g., \texttt{epic\_games} in Fig.~\ref{fig:prompt_pandas}) ; (c) the correct Pandas code template for the task; (d) an NL explanation of the output. Including dataset information is motivated by past work in text-to-SQL~ \cite{bogin-etal-2019-representing,wang-etal-2020-rat}.

Step (d) is influenced by the Chain-of-Thought (CoT) prompting paradigm, which has demonstrated enhanced performance in multi-step tasks \cite{Wei2022ChainOT,Wang2022SelfConsistencyIC,Suzgun2022ChallengingBT,Yoran2023AnsweringQB}. Following the CoT methodology, we incorporate an explanation for each few-shot example. We use \emph{least-to-most prompting}~\cite{zhou2023leasttomost}, in which we provide
the examples at an increasing level of difficulty. Hence, we gradually ``teach'' the LLM fundamental concepts before progressing to more intricate examples. 
Finally, in part (3), we describe the analytical goal along with a sample of the first five rows of the input dataset.

\vspace{2mm}
\noindent\textit{Pandas-to-LDX.} For the Pandas-to-LDX prompt, its structure mirrors the previous prompt, 
i.e., first presenting the Pandas-to-LDX translation task, few-shots examples, etc. This time, we omit the dataset information (2.b) as it is redundant for this simpler task.

\vspace{1mm}
To evaluate our solution, we constructed a new benchmark dataset, publicly available in~\cite{our_github_repository}. The benchmark dataset consists of 182 instances of analysis goals and corresponding \lang{} specifications, as described in Section~\ref{sec:exp_setup}. The full versions of all prompts are provided in~\cite{our_github_repository}.

\section{Experiments}
\label{sec:experiments}

\begin{table*}[t]
\vspace{-2mm}
\small
\centering
\begin{tabular}{|lll|c|}
\hline
\multicolumn{1}{|c|}{\textbf{}} & \multicolumn{1}{l|}{\textbf{\begin{tabular}[c]{@{}c@{}}Exploration Meta Goal\end{tabular}}} & \textbf{Example (concrete) Goal}                                                  & \textbf{\# Ex.} \\ \hline
\multicolumn{1}{|c|}{1}                        & 
\multicolumn{1}{l|}{Identify an uncommon entity}                                                    & $g_1$: “Find an atypical country” (NETFLIX)                                           & 18                                   \\ \hline
\multicolumn{1}{|c|}{2}                        & \multicolumn{1}{l|}{Examine a phenomenon (subset)}                                         & $g_2$: “Examine characteristics of successful TV shows”   (NETFLIX)                                    & 16                                   \\ \hline
\multicolumn{1}{|c|}{3}                        & \multicolumn{1}{l|}{Discover contrasting subsets}                                                    & $g_3$: "Find three actors with contrasting traits” (NETFLIX)                         & 22                                   \\ \hline
\multicolumn{1}{|c|}{4}                        & \multicolumn{1}{l|}{Survey an attribute}                                                & $g_4$: "Survey apps’ price” (PLAY STORE)                                       & 21                                   \\ \hline
\multicolumn{1}{|c|}{5}                        & \multicolumn{1}{l|}{Describe an unusual subset  }                           & $g_5$: "Highlight distinctive characteristics of summer-month flights”   (FIGHTS)                               & 27                                   \\ \hline
\multicolumn{1}{|c|}{6}                        & \multicolumn{1}{l|}{Investigate various aspects of an attribute}                                   & $g_6$: "Investigate reasons for delay” (FLIGHTS)                                          & 22                                   \\ \hline
\multicolumn{1}{|c|}{7}                        & \multicolumn{1}{l|}{Explore through a subset}                                          & $g_7$: "Analyze the dataset, with a focus on flights affected by weather-related delays” (FLIGHTS)    & 28                                   \\ \hline
\multicolumn{1}{|c|}{8}                        & \multicolumn{1}{l|}{Highlight interesting sub-groups}                                             & $g_8$: "Highlight interesting sub-groups of apps with at least 1M installs” (PLAY STORE) & 28                                   \\ \hline

\end{tabular}
\caption{Overview of the Goal-Oriented ADE Benchmark (182 Instances)}
\vspace{-6mm}
\label{tab:tasks}
\end{table*}

We implemented \system{} in Python 3: The  \lang{} verification engine utilizes the Tregex Python implementation in ~\cite{tregeximplementation}, and our CDRL engine is built in  ChainerRL~\cite{fujita2019chainerrl}, based on the DRL framework for data exploration, publicly available in~\cite{atena_repo}. All our experiments were run on a 24-core CPU server.
The full experiments code and data are provided in our Github repository~\cite{our_github_repository}.

Our experiments are conducted along three key facets: (1) Success in deriving correct \lang{} specifications given an analytical goal and dataset; (2) relevance and usefulness of auto-generated exploratory sessions; and (3) performance and ablation study of our CDRL-based modular ADE engine. 

We next describe the construction and properties of our benchmark dataset, then provide the details for each experiments set.

\subsection{Benchmark Dataset for Goal-oriented ADE}
\label{sec:exp_setup}

We constructed the first benchmark dataset, to our knowledge,
for goal-oriented exploration specifications. Our dataset comprises 182 pairs of analytical goals and their corresponding exploration specifications in \lang{} using three different tabular datasets: \textit{(1) Netflix Titles Dataset}~\cite{netflixkaggle} with 9K rows and 11 attributes, 
\textit{(2) Flight-delays Dataset ~\cite{flightskaggle}}, with 5.8M rows and 12 attributes, 
and \textit{(3) Google Play Store Apps ~\cite{playstorekaggle}}, with 10K rows and 11 attributes. 

To build the benchmark dataset, we first characterized 8 exploration ``meta-goals'', as depicted in Table~\ref{tab:tasks}. The meta-goals  were selected by analyzing 36 real-life exploration notebooks available on Kaggle, for the Netflix, Flights, and Playstore datasets (See~\cite{netflixkaggle,flightskaggle,playstorekaggle}), and in accordance with~\cite{wongsuphasawat2019goals} and~\cite{alspaugh2018futzing}, in which the authors identify common analytical tasks, among them several open-ended exploration goals (questions). We then chose an exemplar \textit{concrete} goal for each meta goal (See $g_1$-$g_8$ in Table~\ref{tab:tasks}), and composed \lang{} specifications $Q_X^1-Q_X^8$ based on the content of relevant exploration notebooks on~\cite{netflixkaggle,flightskaggle,playstorekaggle}. $Q_X^1$ is depicted in Figure~\ref{fig:ap_workflow}c, and the rest of the queries are available on~\cite{our_github_repository}. 

\begin{figure}[t]
    \includegraphics[width=0.8\linewidth]{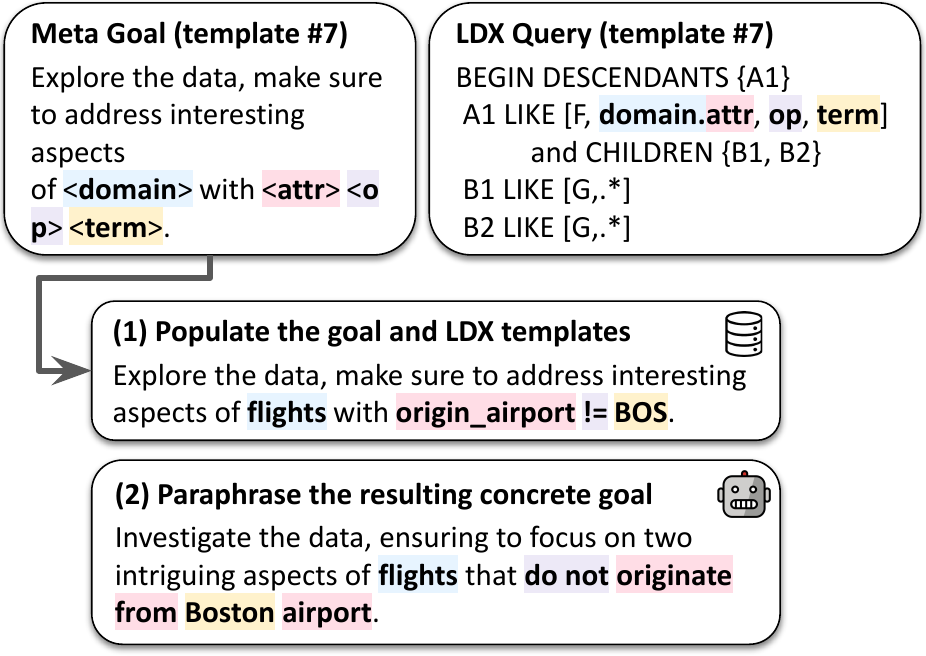}
    \vspace{-3mm}
    \caption{Benchmark Dataset Generation}
    \label{fig:dataset_generation}
\end{figure}

Then, to extend our dataset from $8$ instances to $182$, we adopted the scheme outlined in Figure~\ref{fig:dataset_generation}. First, we stripped each exemplar pair $(g_i,Q_X^i)$ from any dataset-related trait such as attribute names, aggregative operations, and predicates defining data subsets, thus creating ``template'' goal descriptions and \lang{} queries. Next, we populated the goal and \lang{} templates by randomly incorporating values from our three datasets. For instance, the templates in Figure~\ref{fig:dataset_generation}, associated with Meta-Goal 7 (See Table~\ref{tab:tasks}) are populated using the \texttt{Flights}~\cite{flightskaggle} data domain, the \texttt{origin\_airport} attribute, operator $\neq$,and the term \texttt{`BOS'} (the populated \lang{} template is omitted for brevity).
Next, since the populated goal description templates may sound unnatural, we utilized an LLM-based paraphrasing approach (implemented with ChatGPT). This resulted in a more naturally phrased, rich, and diverse set of 200 analytical tasks, out of which we manually discarded 18 nonsensical goals, that did not reflect a realistic user intent. Table~\ref{tab:tasks} lists total number of instances for each meta goal (See~\cite{our_github_repository} for full details).

\subsection{Specifications Derivation Performance}
\label{sec:llm_exp}
We first gauge the effectiveness of our LLM component in deriving correct \lang{} specifications (Full output sessions are evaluated in Section~\ref{sec:exp_us}). 
We analyze the \lang{} derivation performance in four experimental scenarios, varying whether the dataset or meta-goals are seen or unseen in the few-shot prompt examples. We compare the results of our two-stage solution to a single prompt approach that generates \lang{} directly.

\subsubsection*{Experimental Settings}
We now detail the evaluation measures and provide an overview of the different scenarios and baselines.

\noindent\textbf{Evaluation Metrics.}  Evaluating text generation quality is a known challenge with various approaches~\cite{Yaghmazadeh2017SQLizerQS,Yu2018SpiderAL,Kapanipathi2020LeveragingAM}. For example, Text-to-SQL performance assessments often rely on query execution results~\cite{Yaghmazadeh2017SQLizerQS,scholak-etal-2021-picard,rubin-berant-2021-smbop}, but this is unsuitable for \lang{} specifications as they span a multitude of compliant output sessions. Alternative measures include exact string match \cite{Yu2018SpiderAL,acl18sql}, and graph edit distance, commonly used for graph semantic parsing tasks \cite{Cai2013SmatchAE, Kapanipathi2020LeveragingAM}. Drawing inspiration from these, we introduce two measures for comparing the generated \lang{} queries against the gold benchmark queries: the \emph{Two-way Levenshtein Distance}, which considers the query strings, and the \textit{Exploration Tree Edit Distance} from~\cite{reactkdd}, focusing on the parsed output of the queries.

\vspace{1mm}
\textit{(1) Two-way Levenshtein distance ($lev^2$).} Levenshtein distance is commonly used to measure the character overlap between two strings. However, its standard implementation falls short in the context of \lang{}, as two queries may be conceptually similar but differ, for instance, in the order of operations. To address this limitation, we computed the string distance separately for structural and operational specifications, and then aggregated the two scores. The structure score, denoted as $\widebar{lev}(Q_{struct}, Q'_{struct})$, represents the normalized Levenshtein score when omitting operational specifications. The operational distance is defined by $\frac{1}{|Q_{opr}|}*\sum_{o\in Q_{opr}}{min_{o'\in Q'_{opr}}{\widebar{lev}(o,o')}}$. In this expression, $Q_{opr}$ and $Q'_{opr}$ are sets of operational specifications in the two compared \lang{} queries. We sum the distance scores, for each operation $o$ in $Q_{opr}$ of the most similar operation in the compared \lang{} query $Q'_{opr}$, and then divide the result by the size of $Q_{opr}$. The final $lev^2$ is computed as the harmonic mean of the inverses of each score.



\vspace{1mm}
\textit{(2) Exploration Tree Edit Distance ($xTED$).} We employ the exploration tree edit distance, proposed in~\cite{reactkdd}. This measure augments the tree edit distance~\cite{zhang1989simple} function with a dedicated label distance notion to assess the distinction between two  query operations (See~\cite{reactkdd} for full detail). To apply this metric, we construct a minimal tree for each compared \lang{} query while masking the continuity variables (see Appendix~\ref{sec:xted_implementation_details} for more details). 

\vspace{1mm}
Since both $lev^2$ and $xTED$ scores represent normalized distance functions, we consider their complements (i.e., $1 - \textit{score}$), where a higher value is indicative of better performance.

\vspace{1mm}
\noindent\textbf{Scenarios and Baselines.} We conducted four distinct experimental scenarios involving the presence or absence of dataset and meta-goals in the few-shot prompts. In each scenario, the model receives a \textit{test} analytical goal and dataset (selected from the 182 instances in Table~\ref{tab:tasks}) and asked to generate appropriate \lang{} specifications.
In the simplest scenario \textit{(1) seen dataset and meta-goal}, the prompts, as described in Section~\ref{sec:ldx_llm}, include few-shot examples over the same test dataset and meta-goal associated with the test goal (\textit{excluding} the test goal itself). In the subsequent scenarios \textit{(2) seen dataset, unseen meta-goal} and \textit{(3) unseen dataset, seen meta-goal}, prompts include examples from the same dataset, excluding the associated meta-goal, and vice versa.
In the most challenging scenario \textit{(4) unseen dataset and meta-goal}, few-shot examples are provided from different datasets and different meta-goal compared to the test goal. 
Importantly, in \textit{no scenario} the model obtains an example of the exact same analysis goal used in the test.  

To evaluate the efficacy of our NL2PD2LDX solution, we contrasted it with a direct NL2LDX prompt, where the LLM directly generates \lang{} specifications (See Appendix~\ref{app:nl2ldx_direct}). We assessed the performance for both ChatGPT (gpt-3.5-turbo)\cite{gpt35} and GPT-4~\cite{openai2023gpt4} (for both LLMs we used 0 temperature to obtain consistent results).

\paragraph*{Results.} Table~\ref{tab:combined-llm-results} presents the results for both ChatGPT and GPT-4, with and without our chained prompt solution (denoted +PD in the table), across all four scenarios.
First, in the easiest Scenario 1 (\textit{(1) seen dataset and meta-goal}), both LLMs perform well, with GPT-4 achieving optimal results as expected. See that the chained prompt solution exhibits negligible impact, suggesting that the presence of the meta-goal within the prompt allows for easy overfitting, reducing the need for an intermediary solution.
In Scenario 2 (\textit{seen dataset, unseen meta-goal}), the performance of both LLMs decreases as the few-shot examples diverge from the test task. Here, a significant improvement (more than 5 points) is achieved by employing our NL2PD2LDX solution for both models, with GPT-4+PD yielding the best results. 
Moving to Scenario 3 (\textit{unseen dataset, seen meta-goal}), see that the overall performance is better than in Scenario 2, as both LLMs tend to generalize better to unseen datasets than to unseen meta-goals. While our chained solution boosts ChatGPT results by more than 5 points, GPT-4 still achieves the highest score, almost on par with the results in Scenario 1.
Lastly, in the most challenging scenario 4 (\textit{(unseen dataset, unseen meta-goal}), the chained solution yields higher scores for both LLMs, with GPT-4+PD slightly outperforming ChatGPT+PD.

\vspace{1mm}
\noindent\textit{Summary.} Our experimental results show the efficacy of our LLM-based solution, even when the entire class of analytical goals (i.e., all goals with a similar intent) or the dataset are new to the model. We note that further improvements can be achieved, especially for the latter, most challenging scenario where both the datasets and the meta-goals are unseen. We therefore make our benchmark dataset public~\cite{our_github_repository}, to facilitate the evaluation of future solutions.

\begin{table}[t]
\small
\centering

\begin{tabular}{|lcccc|}
\hline
\multicolumn{1}{|l|}{\multirow{2}{*}{\textbf{Model\textbackslash{}Settings}}} & \multicolumn{2}{c|}{\textbf{Seen Meta-Goal}}                                        & \multicolumn{2}{c|}{\textbf{Unseen Meta-Goal}}                 \\ \cline{2-5} 
\multicolumn{1}{|l|}{}                                                        & \multicolumn{1}{c|}{$\bm{lev^2}$}       & \multicolumn{1}{c|}{$\bm{xTed}$}       & \multicolumn{1}{c|}{$\bm{lev^2}$}       & $\bm{xTed}$       \\ \hline
\multicolumn{5}{|c|}{\textbf{Seen Dataset}}                                                                                                                                                                                         \\ \hline
\multicolumn{1}{|l|}{\textbf{ChatGPT}}                                        & \multicolumn{1}{c|}{0.87}                & \multicolumn{1}{c|}{0.87}                & \multicolumn{1}{c|}{0.64}                & 0.6                 \\ \hline
\multicolumn{1}{|l|}{\textbf{ChatGPT +   Pd}}                                 & \multicolumn{1}{c|}{0.89}                & \multicolumn{1}{c|}{0.89}                & \multicolumn{1}{c|}{0.72}                & 0.69                \\ \hline
\multicolumn{1}{|l|}{\textbf{GPT-4}}                                          & \multicolumn{1}{c|}{{\underline{\textbf{0.97}}}} & \multicolumn{1}{c|}{{\underline{\textbf{0.97}}}} & \multicolumn{1}{c|}{0.71}                & 0.7                 \\ \hline
\multicolumn{1}{|l|}{\textbf{GPT-4 + Pd}}                                     & \multicolumn{1}{c|}{0.97}                & \multicolumn{1}{c|}{0.96}                & \multicolumn{1}{c|}{{\underline{ \textbf{0.77}}}} & {\underline{\textbf{0.75}}} \\ \hline
\multicolumn{5}{|c|}{\textbf{Unseen Dataset}}                                                                                                                                                                                       \\ \hline
\multicolumn{1}{|l|}{\textbf{ChatGPT}}                                        & \multicolumn{1}{c|}{0.79}                & \multicolumn{1}{c|}{0.79}                & \multicolumn{1}{c|}{0.65}                & 0.65                \\ \hline
\multicolumn{1}{|l|}{\textbf{ChatGPT +   Pd}}                                 & \multicolumn{1}{c|}{0.86}                & \multicolumn{1}{c|}{0.84}                & \multicolumn{1}{c|}{0.72}                & 0.68                \\ \hline
\multicolumn{1}{|l|}{\textbf{GPT-4}}                                          & \multicolumn{1}{c|}{{\underline{\textbf{0.95}}}} & \multicolumn{1}{c|}{{\underline{\textbf{0.95}}}} & \multicolumn{1}{c|}{0.71}                & 0.7                 \\ \hline
\multicolumn{1}{|l|}{\textbf{GPT-4 + Pd}}                                     & \multicolumn{1}{c|}{0.94}                & \multicolumn{1}{c|}{0.93}                & \multicolumn{1}{c|}{{\underline{\textbf{0.73}}}} & {\underline{\textbf{0.71}}} \\ \hline
\end{tabular}\caption{Specification Derivation (NL-to-LDX) Results}
\label{tab:combined-llm-results}
\end{table}

\begin{figure*}[t]
\vspace{-1mm}
\centering
\begin{minipage}[t]{0.68\textwidth}
      \centering
      \includegraphics[width=1\textwidth]{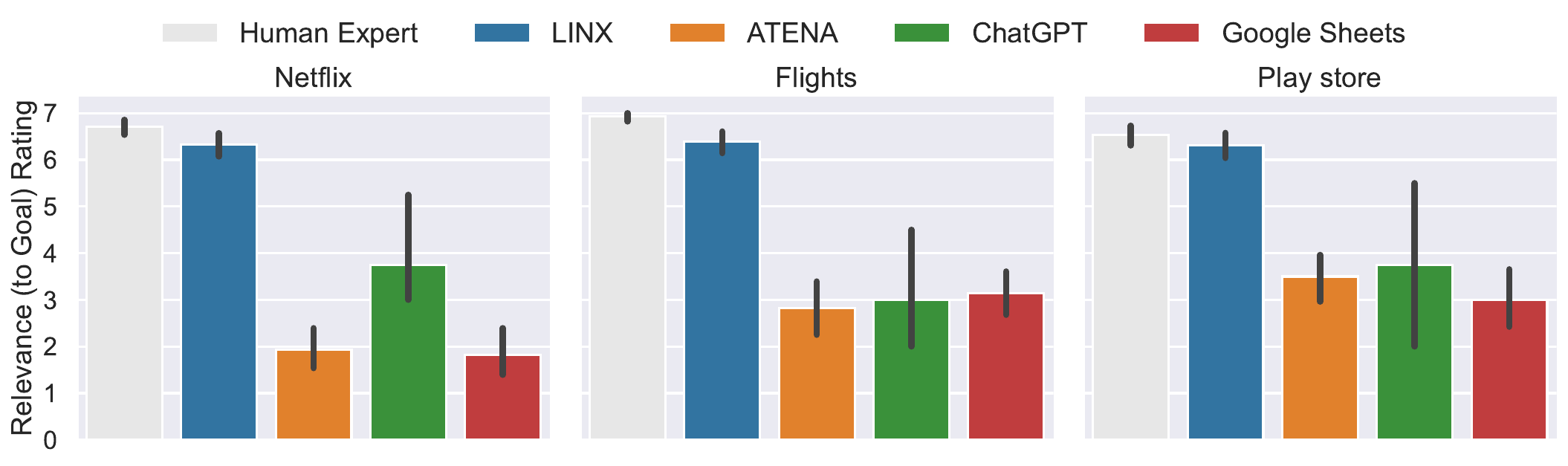}
    \vspace{-8mm}
    \caption{User Study -- Relevance Rating of Exploration Notebooks to the Given Goal}
    \label{fig:UserStudyRelevance}
   \end{minipage}
 \quad\quad  
\begin{minipage}[t]{0.25\textwidth}
    \centering
    \includegraphics[width=1\textwidth]{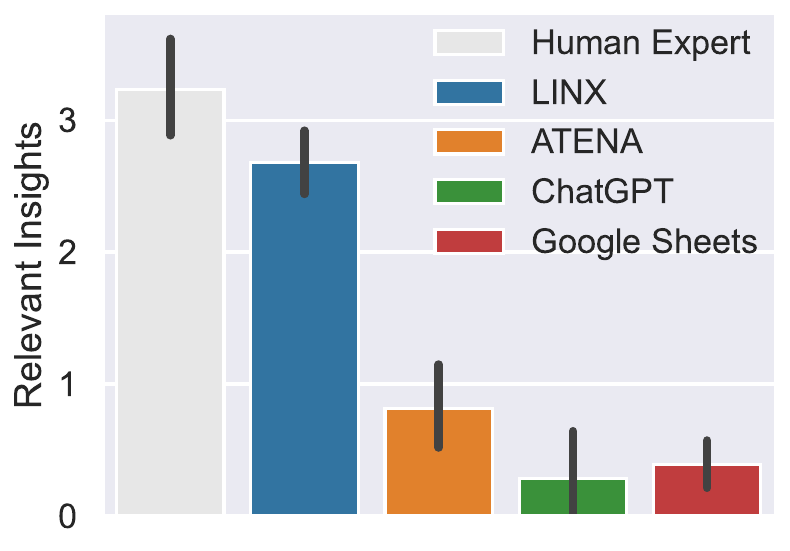}
    \vspace{-8mm}
    \caption{Avg. Num. Insights}
    \label{fig:UserStudyInsights}
    \end{minipage}
\end{figure*}

\subsection{Relevance and Quality (User Study)}
\label{sec:exp_us}

We next evaluate the overall quality and relevance of exploration sessions generated by \system{}, compared to sessions generated by alternative baselines. We conducted both a \textit{subjective} study, were users are asked to rate the output sessions according to numerous criteria, and an \textit{objective} study, where we measured users' performance in inferring relevant insights w.r.t. the analytical goal. 

\paragraph*{Experiment Setup}
We recruited a total of 30 participants, by publishing a call for CS students or graduates that are familiar with data analysis yet are not subject matter experts. We then selected 12 analysis goals and \lang{} specifications from our benchmark dataset.
We used $g_1$-$g-8$, as depicted in Table~\ref{tab:tasks}, and four additional pairs (deferred to~\cite{our_github_repository} for space constraints), to obtain a total of four different goals for each of our three datasets. 


We used \system{} to generate an exploration notebook for each goal and dataset, and presented the output session in a Jupyter notebook (see Figure~\ref{fig:ap_workflow}e for a snippet, and the full notebooks in~\cite{our_github_repository}). 
We evaluated \system{} compared to the the following baselines: \textbf{(1) ATENA~\cite{atena_short}.} We ran ATENA on each of the datasets. As it automatically generates an exploration session but does not accommodate user specifications, it produces the same exploration notebook for all four tasks of each dataset. \textbf{(2) ChatGPT (gpt-3.5-turbo)~\cite{gpt35}}. In this baseline we generated notebooks by asking the LLM to directly build an entire exploration notebook, containing real Pandas code, for a given description of the dataset and an analytical task. We executed the code provided by the LLM and presented the results in a Jupyter notebook. \textbf{(3) Google Sheets Explorer~\cite{googlesheet}.} A commercial ML-based exploration tool that accommodates limited user specifications, allowing to specify columns and data subsets of interest. The specifications were composed w.r.t. to the \lang{} queries for each goal. For example, for goal $g5$ (``characteristics of summer flights''), we selected the columns `month', `airline', `delay-reason' and `scheduled arrival'/`departure', and the data subset containing flights from July and August.
\textbf{(4) Human Expert.} Last, we used exploration notebooks generated manually by experts data scientists, to provide an ``upper bound'' for the output quality of the automatic approaches. We asked three experienced data scientists to manually compose a notebook (\textit{without} any assistive tool) of interesting query operations that are relevant for the given goal. 
The instructions, data, and output of all baselines are provided in our code repository~\cite{our_github_repository}.




\begin{figure}[t]
    \vspace{-1mm}
        \includegraphics[width=0.9\linewidth]{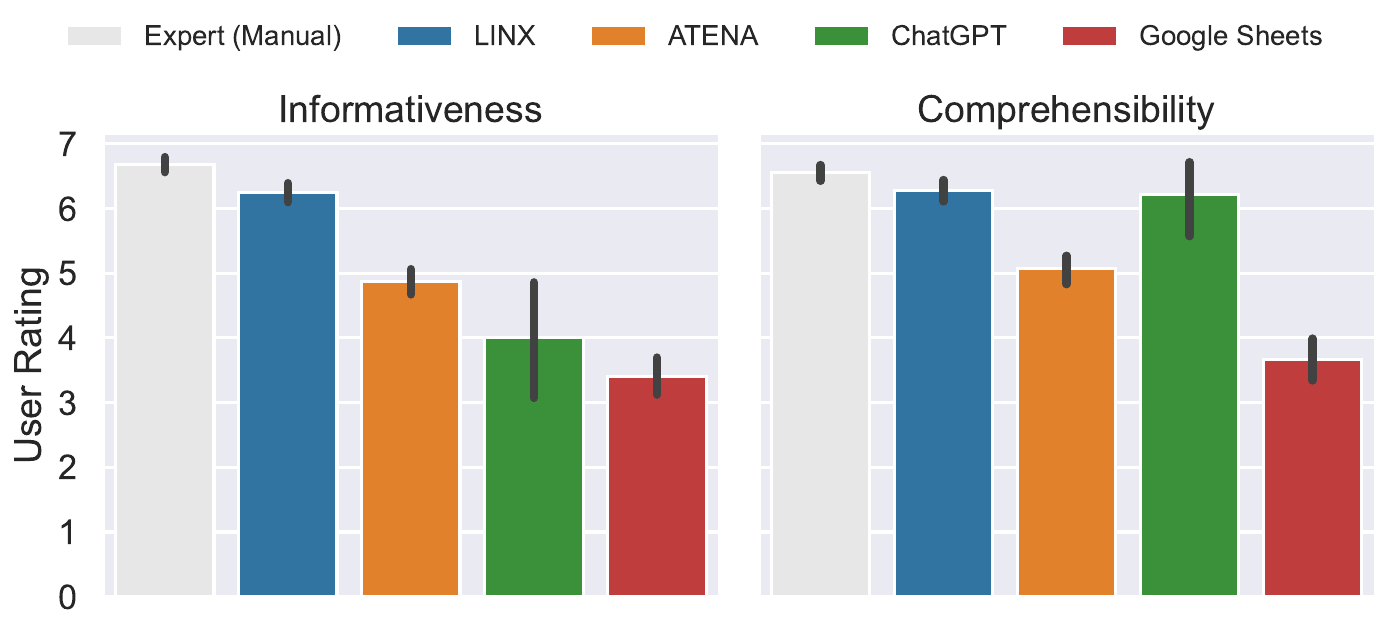}
        \vspace{-3mm}
        \caption{Informativeness \& Comprehensibility Rating}
        \label{fig:us}
    \end{figure}

\paragraph*{Subjective Study (User Rating)}
In this study, the participants were asked to review notebooks, generated by either \system{} or the baselines, w.r.t. each notebook's corresponding analytical task. Each participant reviewed one notebook per dataset to neutralize the effect of experience. We then asked the participants to rate each notebook 
on a scale from 1 (lowest) to 7 (highest) according to the following criteria: 
(1) \textit{Relevance} - \textit{To what degree is the exploration notebook relevant for the given analysis goal?} 
(2) \textit{Informativeness} —  \textit{To what extent does the notebook provide useful information about the data?}
(3) \textit{Comprehensibility} - \textit{To what degree is the notebook comprehensible and easy to follow?}



\vspace{1mm}
 Figure \ref{fig:UserStudyRelevance} presents the \textit{relevance score} of \system{} and the baselines for each of the three datasets. The results are averaged across all participants and goals for each dataset (The vertical line depicts the .95 confidence interval.)
As expected, manually composed notebooks from human experts obtained the highest rating - 6.71, 6.92, 6.53 for the Netflix, Flights, and Play Store datasets (resp).
However, see that \system{} obtains very close scores -- 6.32, 6.39, 6.30 (resp.) for its automatically generated sessions. 

Next, see that the relevance ratings of ChatGPT are lower (3 to 3.75). While ChatGPT does support natural language specifications, it mainly outputs descriptive statistics and simple aggregations. For example, for goal $g_1$ on the Netflix dataset, ChatGPT generates pandas code (deferred to our Github repository in~\cite{our_github_repository}) alongside the following description of its logic: \textit{We first filter the dataset to separate TV shows and movies. Then, we calculate the global percentage of TV shows and movies by dividing the number of TV shows/movies by the total number of entries in the dataset. 
Next, we group the dataset by country and calculate the percentage of TV shows and movies for each country.
We then compare the country percentages to the global percentages and identify countries with a difference of more than 10\%.} 

See that ChatGPT defines an atypical country as one that has a 10\% deviation in the ratio of movies to TV-shows, which results in multiple countries that the user than needs to manually examine. 


Finally, The scores of ATENA and Google Sheets are lower, reaching about 2-3 out of 7. Naturally, the fact that ATENA does not support user specifications, and Google Sheets supports only limited specifications  -- makes their solutions insufficient for generating \textit{relevant} notebooks for the given analytical goals. 

\vspace{1mm}
Next, we inspected the \textit{informativeness} and \textit{comprehensiveness} scores.
Figure \ref{fig:us} depicts the average scores, over all three datasets. (The black vertical lines represent the .95 confidence interval.)

The human-expert notebooks once again achieve the highest scores. Additionally, both ATENA and Google Sheets now attain higher scores: ATENA scores 4.86 and 5.07, while Google Sheets follows with 3.40 and 3.67 for informativeness and comprehensiveness, respectively. ChatGPT achieves a high comprehensiveness score of 6.21, primarily due to its utilization of very simple analytical operations and straightforward code documentation. However, it falls behind in terms of informativeness, scoring an average of 4/7.


\begin{table}[t]
\centering
\small
\begin{tabular}{ |p{8cm}| } 
      \hline
    \textbf{User Insight (for goal $g_i$)} \\ [0.5ex] 
    \hline
    ``The ratio of movies-series in India is higher than the movies-series ratio anywhere else.'' ($g_1$) \\ 
    \hline
    ``Most multi-season US TV shows are dramas or comedies'' ($g_2$) \\ 
    \hline
    ``About one-third of flights occur in summer, yet the monthly rate of delays remains consistent throughout the year.'' ($g_5$) \\ 
    \hline
    ``While long flights are not delayed often, if they are, this is mainly for a security reason.'' ($g_6$)\\ 
    \hline
    ``Apps with 1M installs are typically free, highly rated, and compatible with Android 4.'' ($g_8$)\\
    \hline
\end{tabular}

\caption{Examples of Insights Derived by Users Using \system{}}
\label{tab:insights}
\vspace{-3mm}
\end{table}

Interestingly, \system{} still obtains higher scores than ATENA, ChatGPT, and Google Sheets, (6.24 and 6.28 for informativeness and coherency). This particularly shows that \system{} does not compromise on informativeness or comprehensibility, when generating \textit{goal-oriented} exploratory sessions. 


\subsubsection*{Objective Study (Task Completion Success)}


We also compared \system{} with the baselines in an objective manner -- by asking users to examine notebook and then extract a list of insights that are \textit{relevant} w.r.t. the given analytical goal.
The correctness and relevance of insights were evaluated, by the same experts who constructed the manual \textit{human-expert} notebook (Baseline 4), and is therefore highly familiar with the datasets and respective goals.

Figure \ref{fig:UserStudyInsights} shows the average number of goal-relevant insights derived using each baseline. 
Using \system{}, users derived an average of 2.7 relevant insights per goal, which is second only to the human-expert notebooks (3.2 insights).
ATENA and Google Sheets are again far behind with an average of 0.8 and 0.4 relevant insights per goal (resp). Interestingly, ChatGPT obtains the lowest score of 0.3 insights. This is because for the vast majority of tasks, users could not derive any explicit insight (since, as mentioned above, ChatGPT notebooks contained mostly general descriptive statistics).

Last, to further examine the quality of the insights derived from \system{}-generated notebooks,
we provide example insights derived by the participants, depicted in Table~\ref{tab:insights}. See that users were able to extract compound, non-trivial insights that are indeed relevant to the corresponding analytical goals.

\vspace{2mm}
\noindent\textit{Summary.} An extensive user study with 30 participants shows that users not only rate the exploration notebooks generated by \system{} as highly relevant, informative, and comprehensible, but were also able to derive significantly more relevant insights compared to the non-human baselines.

\subsection{CDRL Performance \& Ablation Study}
\label{sec:exp_performance}
Last, we examine the performance of our CDRL Engine, by conducting first an ablation study, then a convergence comparison with the goal-invariant ATENA~\cite{atena_short} ADE system.

\paragraph*{Ablation Study} 
To gauge the necessity in the components of \system{} we compared it to the following system versions, each missing one or more  components:
\textbf{(1) \basebin{}} uses a binary end-of-session reward, based solely on the output of the \lang{} verification engine, without using our full reward scheme (§\ref{sec:reward})  and specification-aware network (§\ref{sec:network_arch}). Instead, it uses the basic neural network of~\cite{atena_short}.  \textbf{(2) \basedist{}} uses the reward scheme, as described in Section~\ref{sec:reward},
without the immediate reward and the specification-aware network. 
\textbf{(3) \baseimm{}} uses the full reward scheme (including the immediate reward), but with the basic neural network of~\cite{atena_short}.


\vspace{1mm}
We employed each baseline on the same 12 \lang{} queries used in the user study, and examined how many of the generated exploration sessions were indeed compliant with the input specifications.
The results are depicted in Table~\ref{tab:AblationStudyResults}, reporting the baselines' success in: (1) \textit{structural} compliance, where a generated notebook complies with the structural specifications but not the operational ones, and (2) \textit{full compliance}, where all specifications are met.  

First, see that \textit{\basebin{}}, which only receives the binary, end-of-session reward, fails to generate compliant sessions for any of the queries. As mentioned above, this is expected due to the sparsity of the reward and the vast size of the action space. 
\textit{\basedist{}}, which uses the more flexible compliance reward at the end of each session. obtains better results -- fully complying with 3 queries, and structure-compliant with 7 additional ones. 
Next, \textit{\baseimm{}} obtains a significant improvement -- it is able to comply with the structural specifications of all 12 \lang{} queries. However, it was \textit{fully} compliant only for 5/12 queries. 

Finally, see that only the full version of \system{}-CDRL, which uses both the full reward-scheme \textit{and} the specification-aware neural network, is able to generate compliant sessions for 100\% of the \lang{} queries. 
This shows that our adaptive network design, as described in Section~\ref{sec:network_arch}, is particularly useful in encouraging the agent to perform specification-compliant operations -- despite the inherently large size of the action-space.

\begin{figure}[t]
\vspace{-9mm}
    \includegraphics[width=0.99\linewidth]{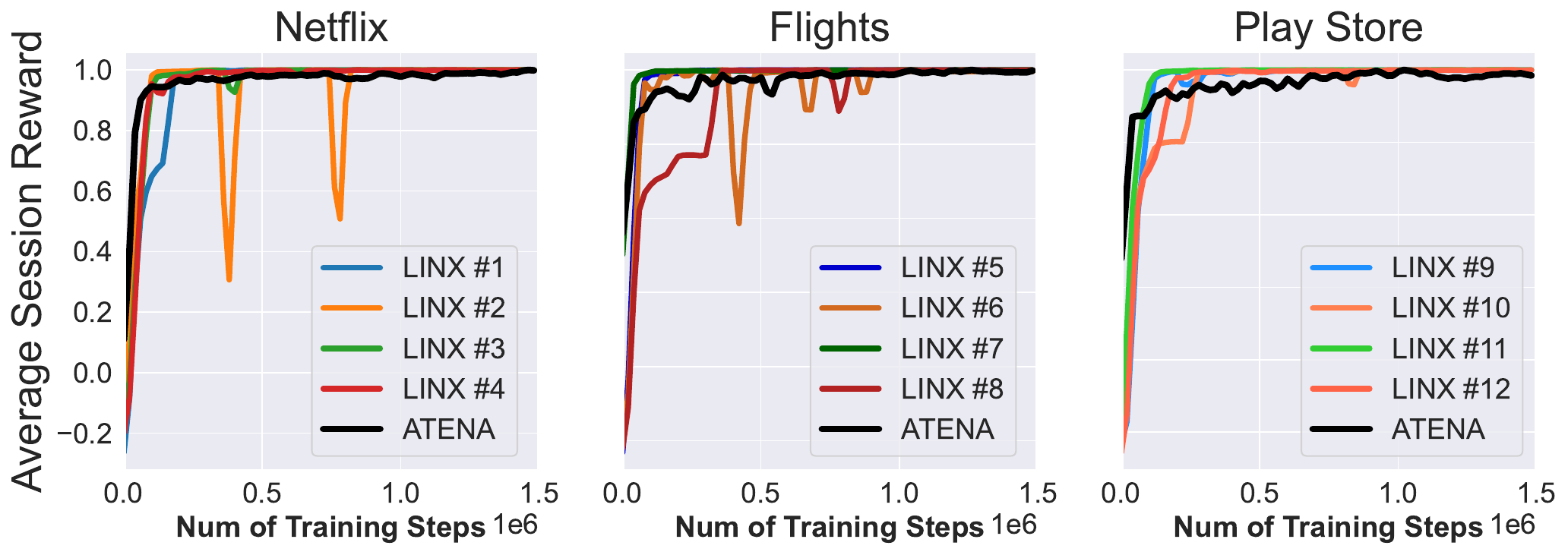}
    \vspace{-3mm}
    \caption{Convergence Comparison to ATENA}
    \vspace{2mm}
    \label{fig:ConvergancePlot}
\end{figure}

\paragraph*{Convergence Performance}
Lastly, we examine the performance of our CDRL engine and compare it with the DRL engine of ATENA~\cite{atena_short}. Figure~\ref{fig:ConvergancePlot} shows the convergence plots for the 12 \lang{} queries. The convergence for each \lang{} query $i$ (corresponding to goal $g_i$) is depicted using a line labeled \textit{`\system{} \#$i$'}, with the black line in each figure representing the convergence process of ATENA~\cite{atena_short}, serving as a baseline. (Recall that ATENA can only produce one generic exploration session per dataset.) Since the maximal reward varies depending on the \lang{} query and dataset, we normalize the rewards so that the maximum is 100\%. Observe that the convergence processes of both ATENA and \system{}-CDRL are roughly similar. Notably, \system{}-CDRL sometimes converges even faster than ATENA (e.g., for the Play Store dataset), where ATENA takes 0.85M steps and \system{} only 0.4M steps on average. In general, the average convergence to 100\% reward is 0.36M steps, which takes about 20 minutes on our simple CPU hardware. While this could be significantly improved with GPU hardware, it is important to note that such running times are acceptable as \system{} is \textit{not} intended for interactive analysis (See again our discussion in Section~\ref{sec:prelim}).

\begin{table}[t]
\small
\centering
\begin{tabular}{ |c|c|c| } 
    \hline
    
    \textbf{\system{} Version} & \textbf{Structure Compliance}& \textbf{Full Compliance}  \\  
    \hline
    Binary Reward Only & 0/12 (0\%) & 0/12 (0\%) \\ 
    \hline
    Binary+Imm. Reward & 10/12 (84\%) & 3/12 (25\%) \\ 
    \hline
    W/O Spec. Aware NN & 12/12 (100\%) & 5/12 (42\%) \\ 
    \hline
    \textbf{\system{}-CDRL (Full)} & \textbf{12/12 (100\%)} & \textbf{12/12 (100\%)} \\ 
    \hline
\end{tabular}
\caption{Ablation Study Results}
\vspace{1mm}
\label{tab:AblationStudyResults}
\end{table}

\vspace{1mm}
\noindent\textit{Summary.} Our performance study demonstrates two key findings: (1) all elements of \system{}-CDRL, are required for consistently generating compliant notebooks. (2) Despite its more complex reward system and neural network, the convergence performance of \system{} is on par with ATENA.

    \section{Conclusion \& Future Work}
\label{sec:conclusion}

This paper introduces \system{}, a generative system for automated, goal-oriented exploration. Given an analytical goal and a dataset, \system{} combines an LLM-based solution for deriving exploratory specifications and a modular ADE engine that takes the custom specifications and generates a personalized exploratory session in accordance with the input goal.


In future work, we will explore ways in which LLMs can further enhance the analytical process. A promising direction is to utilize LLMs for augmenting \system{} notebooks with additional elements like captions, explanations, and visualizations, while also considering auto-visualization solutions such as~\cite{wongsuphasawat2017voyager,lee2021lux}. Another direction is the adaptation of \system{} to interactive analysis and data manipulation code generation. 


\clearpage
\bibliographystyle{ACM-Reference-Format}
\bibliography{bib2}
\clearpage
\appendix

\section*{APPENDIX}

\section{Additional Material \& Discussions}

\subsection{\lang{} Comparison to Tregex}

\noindent As mentioned above, \lang{} is based on Tregex~\cite{levy2006tregex}, a popular tree query language among the NLP community. Tregex is typically used to query syntax trees, grammars and annotated sentences. \lang{} adopts similar syntax for specifying the exploratory tree structure and labels (exploratory operations), and extends it to the context of data exploration using the \textit{continuity variables}, which semantically connect the desired operations as described above.

\subsection{ \lang{} Verification Engine Computation Times Discussion}

As is the case for Tregex~\cite{skala2014structural}, 
evaluating an \lang{} query may require, in the worst case, iterating over all possible node assignments, of size  $\frac{|\stree|!}{(|\stree|-|\nnodes|)!}$.
However, we show in Section~\ref{sec:exp_performance} that in practice, computing the \lang{}-compliance reward scheme takes a negligible amount of time, compared to the overall session generation time. 
This is mainly because both the exploration tree $\stree$ and the query $\lquery$ are rather small\footnote{For example, the mean session size in the exploratory sessions collection of~\cite{reactkdd} is 8.}.

\subsection{\lang{}-Compliance Reward Scheme}
\label{app:reward}
The generic exploration reward, as described above, encourages the agent to employ interesting, useful analytical operations on the given dataset. However, the session is inadequate if it is interesting yet irrelevant to the goal at hand.  Our goal is therefore to enforce that the agent also produces a sequence of operations that is fully compliant with the \lang{} specifications derived from the analytical goal description.

Given the input dataset and $\lquery$ specifications, our compliance reward scheme gradually ``teaches'' the agent to converge to a $\lquery$-compliant exploratory session. This is based on the observation that \textit{structural specifications should be learned first}. 
Namely, if the agent had learned to generate \textit{correct} operations in an \textit{incorrect} order/structure -- then the learning process is largely futile (because the agent had learned to perform an incorrect exploration path, and now needs to relearn, from scratch, to employ the desired operations in the correct order). 
\system{} therefore encourages the agent to first generate exploratory sessions with the correct structure, using a high penalty for non-compliant sessions.
Once the agent has learned the correct structure, it will now obtain a gradual reward that encourages it to satisfy the operational specifications. 

Finally, when the agent manages to generate a fully compliant session it obtains a high positive reward, and can now further increase it by optimizing on the exploration reward signal (as described above).


\SetCommentSty{rmfamily}
\setlength{\textfloatsep}{2pt}
 \IncMargin{1em}
\begin{algorithm}[t]
\small
\DontPrintSemicolon
\KwIn{Exploration Tree $\stree$, \lang{} Specifications $\slist$}
   \KwOut{Reward $R$}
    
   \SetNlSkip{2mm}
    
    \If{\label{ln:rpos1} $\textbf{Verify\lang{}}(\slist,\stree$) = True}{
     \tcp{$\stree$ is compliant with $\slist$}
    \Return \textit{POS\_REWARD} \label{ln:rpo2} }
    \vspace{1mm}
    $S_{struct} \leftarrow struct(\lquery)$ \tcp{Structural specs of $\slist$} \label{ln:get_structs}
    
    \label{ln:get_all_A}$\Phi_V \leftarrow 
    GetTregexNodeAssg(S_{struct},\stree)$ 
    

    \If{$\Phi_V = \emptyset$}{
    \tcp{$\stree$ violates Structural specs}  
     \Return \textit{NEG\_REWARD} \label{ln:rneg}}
    \vspace{1mm}
    \tcp{Calculate operational-based reward:}  
    
    $S_{opr} \leftarrow opr(\lquery)$ \tcp{Operational specs of $\slist$}
    

   \Return{$\max\limits_{\phi_V \in \Phi_V}$ \textbf{\textit{GetOprReward}}($\phi_V,S_{opr}$)}
    
    
    

  \vspace{2mm}
 \SetKwProg{myfunc}{GetOprReward}{}{}
 \nonl \myfunc{$(\phi_V,S_{opr})$}{
  
$reward \leftarrow 0$
  \label{ln:opr_begin}
  
  \label{ln:opr_for} \For{$s \in S_{opr}$}{
    
    $v_s = \phi_V(Node(s))$  
    $reward \pluseq \frac{\text{\# of matching opr. params in } v_s }{\text{\# of specified params in  } s }$ \label{ln:ratio}
    
    }
   \Return $reward$
   \label{ln:opr_return}

   }

\caption{End-of-Session Conditional Reward}
 \label{alg:session_reward}
\end{algorithm}

We next describe our reward scheme, comprising both an \textit{end-of-session} and an \textit{immediate} reward signals. 

\vspace{1mm}
\noindent\textit{End-of-Session Compliance Reward}.
Our End-of-Session reward scheme is depicted in Algorithm~\ref{alg:session_reward}.
Given a session $\stree$ performed by the agent and the set of specifications $\slist$ (from the \lang{} query $\lquery$),
we first check (Line~\ref{ln:rpos1}) whether the session $\stree$ is compliant with $\lang{}$ using Algorithm~\ref{alg:verification}.
In case $\stree$ is compliant, a high positive reward is given to the agent (Line~\ref{ln:rpo2}).

Next, in case $\stree$ is not compliant with $\lquery$, we now check whether it is compliant with $struct(\lquery)$, the \textit{structural specifications} in $\lquery$ (denote $S_{struct}$ here for brevity) This is done by calling the Tregex engine (since there is no need to verify the continuity variables), which returns all valid assignments $\Phi_V$ for $S_{struct}$ over $\stree$ (Line~\ref{ln:get_all_A}).
If there are no valid assignments, it means that $\stree$ is non-compliant with the specified structure, therefore a fixed negative penalty is returned (Line~\ref{ln:rneg}).

In case $\stree$ satisfies the structural specifications (but not the operational/continuity) -- we wish to provide a non-negative reward that is proportional to the number of satisfied \textit{operational} specifications (and parts thereof). 
Namely, the more specified operational parameters
(e.g., attribute name, aggregation function, etc.)
are satisfied -- the higher the reward.
To do so, we compute the operational reward for each node assignment $\phi_v \in \Phi_V$, returning the maximal one.
The operational reward is calculated in \textit{GetOprReward)} (Lines~\ref{ln:opr_begin}-\ref{ln:opr_return}).
We initialize the reward with $0$, then iterate over each operational specification $s \in S_{opr}$ (Line~\ref{ln:opr_for}), and first, retrieve its assigned node $v_s$ (according to $\phi_V$).
We then compute the ratio of matched individual operational parameters in $v_s$, out of all operational parameters specified in $s$ (Line~\ref{ln:ratio}). This ratio is accumulated for all specifications in $S_{opr}$,
s.t. the higher the number of matched operational parameters, the higher the reward.

\vspace{1mm}
\noindent\textit{Immediate (per-operation) Compliance Reward.} To further encourage the agent to comply with the structural constraints, we develop an additional, \textit{immediate} reward signal, granted after each operation. The goal of the immediate, per-operation reward is to detect, in real-time, an operation performed by the DRL agent that violates the structural specifications.

The procedure is intuitively similar to the \lang{} verification routine (Algorithm~\ref{alg:verification}), yet rather than taking a full session as input it takes an \textit{ongoing} session $\stree^i$, i.e., after $i$ steps, and the remaining number of steps $N-i$. It then assesses whether there is a \textit{future} assignment, in up to $n$ more steps, that can satisfy the structural constraints $struct(\lquery)$. 
This is simply done by calling the Tregex match function \textit{GetTregexNodeAssg}$(struct(\lquery),\stree^{\star})$, with each possible \textit{tree completion} (denoted $\stree^\star$) for the ongoing exploration tree. The completion of the ongoing exploration tree $\stree^i$ simply extends it with $N-i$ additional ``blank'' nodes. 
Starting from the current node $v_i$, blank nodes can be added only in a manner that respects the order of query operations execution in the session (captured by the nodes' pre-order traversal order~\cite{reactkdd}). Namely, each added node $v_j$, s.t. $i < j\leq N$ can be added as an immediate child of $v_{j-1}$ or one of its ancestors). We explain below that the number of possible tree completions throughout a session of size $N$ (including the root) is bounded by $C_{N}$, where $C_N$ is the Catalan number.

In more details, the immediate reward procedure at step $i$ of an ongoing session, has $N-i$ remaining nodes to complete a full possible session tree. In order to bound the number of possible trees at each iteration, we will examine the iteration with the largest number of completions, which is right after the first step of the agent, namely when $i=1$ and $T_i$ is a tree with a root and one child $v_1$ which is the current node.
In this scenario, after adding an additional node $v_2$, we got two possible trees: one where $v_2$ is a child of $v_1$, and another one where $v_2$ is the right sibling of $v_1$. When adding one more node, $v_3$, we have total of 5 possible trees: If $v_2$ is child of $v_1$, then $v_3$ can be a child of $v_2$, right sibling of $v_2$ or right sibling of $v_1$. Otherwise it can be a child of $v_2$ or right sibling of $v_2$. The number of possible trees continue to grow in each iteration. Even though the procedure is iterative, each possible final tree of size $N$ can only be generated once, due to the pre-order traversal manner. Thus, the number of possible trees is bounded by the number of ordered trees of size $N$, which is bounded by $C_{N} = \frac{1}{n+1} {2n \choose n}$, where $C_N$ is the Catalan number~\cite{dershowitz1980enumerations}.
To further improve the procedure performance, we employ the immediate reward only after $i\geq 3$ steps. This way still encourage the agent to comply with the structural constraints, but avoid the large number of tree-completions in the first steps. For example, when $10 \leq N \leq 20$, the number of completions is reduced by 4-5X.

\subsection{Specification-Aware Network - An Example}
\label{ex:network}

Recall that Figure~\ref{fig:ButtonsArchitecture} depicts our specification-aware neural network architecture, deriving its structure from the \lang{} specifications $\lquery$.
We next give an example for an action selection process in accordance with our example workflow depicted in Figure~\ref{fig:ap_workflow}.

Figure~\ref{fig:ButtonsArchitecture} displays an example state in the network, in which the action  probabilities are already computed (colored in light orange). 
First, the agent samples an operation type using the multi-softmax segment $\sigma_{ops}$. As in Figure~\ref{fig:ButtonsArchitecture}, 
`Snippet' obtained the highest probability of 0.5. 
If indeed selected, the specific snippet is chosen using Segment $\sigma_{snp}$. See that the Filter snippet \e{`F, Country, eq, *`} obtains the highest probability (0.6). Now, as the set of free parameters only contains the filter `term' parameter, the agent now chooses a term using Segment $\sigma_3$. The chosen term is `India'.

\section{Additional Implementation Details}

\subsection{Prompts Description}
\label{app:nl2ldx_direct}

\begin{figure}[t]
\vspace{-2mm}
    \includegraphics[width=0.98\linewidth]{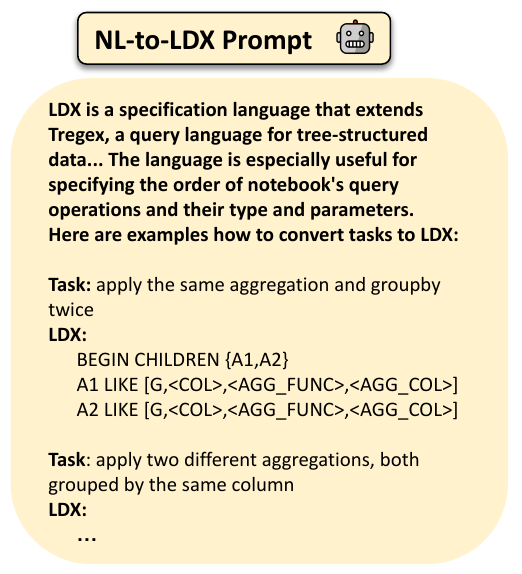}
    \vspace{-3mm}
    \caption{Example of the prompt used for NL2\lang{}}
    \label{fig:prompt_nl2ldx}
\end{figure}

\noindent\textit{NL-to-\lang{}.} The NL-to-\lang{} prompt is structured the same as the NL-to-Pandas prompt as discussed in section §\ref{sec:ldx_llm}: (1) NL-to-\lang{} goal description; (2) a series of few-shot examples; (3) the test analysis goal alongside a small dataset sample.
Each few-shot example in (2) comprises several steps: (a) example goal description; (b) description of the example's corresponding dataset and schema; (c) the correct \lang{} query for the goal; (d) an NL explanation of the output. See Fig.~\ref{fig:prompt_nl2ldx}.

\noindent\textit{Prompts Number of Examples.} The NL-to-Pandas and Pandas-to-\lang{} prompts have 14 and 10 examples respectively, while the NL-to-\lang{} prompt has 14 examples. All prompts contain 8 examples which correspond to our 8 analytical meta-goals (Table~\ref{tab:tasks}). For NL-to-Pandas and NL-to-LDX we added 6 initial examples of mapping basic constructs before moving on to the 8 template examples, and for Pandas-to-\lang{} we added 2 additional general examples.

\subsection{Evaluation Implementation Details}
\label{sec:xted_implementation_details}

\noindent\textit{Minimal Tree.} In section §\ref{sec:llm_exp},it was previously mentioned that we construct a minimal tree for the compared \lang{} queries in order to apply them a tree distance metric. The conversion of \lang{} query to tree is generally straightforward, except for the 'DESCENDANTS' structural specification operator, since the distinction between DESCENDANTS and CHILDREN can't be expressed out-of-the-box. Our ad hoc approach for addressing that is setting the descendants as direct children in the converted tree (meaning the minimal specification-compliant tree) and adding 'children type' as an additional property of the action label. Additionally, we slightly modified the action distance function by penalizing variations in the 'children type' of the compared actions.

\noindent\textit{Masking Continuity Variables.} Furthermore, we mask the continuity variable names of the derived  minimal trees to eliminate prevent scoring reduction due to naming differences. We do so by separately masking each category of continuity variables. For instance, continuity variables that define attributes of filter operations are substituted with identifiers such as att1, att2, att3, and so forth.
Similarly, those that define aggregation function are substituted with identifiers such as aggfunc1, aggfunc2, aggfunc3, etc. This masking systematic approach aids in avoiding false penalty, while on the other hand ensuring scores are not inappropriately inflated.

\end{document}